\documentclass[letterpaper,twocolumn,10pt]{article}
\usepackage{usenix}

\usepackage{url}
\usepackage{hyperref}
\usepackage{paralist}

\usepackage{tikz}
\usepackage{amsmath}

\usepackage{enumitem}

\usepackage{comment}
\usepackage{filecontents}
\usepackage{amsmath}    
\usepackage{booktabs}   
\usepackage{graphicx}   
\usepackage{array}      
\usepackage{xcolor}

\newcommand{\commentone}[1]{{\color{black}#1}}
\newcommand{\commenttwo}[1]{{\color{black}#1}}
\newcommand{\commentthree}[1]{{\color{black}#1}}
\newcommand{\commentfour}[1]{{\color{black}#1}}
\newcommand{\commentfive}[1]{{\color{black}#1}}
\newcommand{\commentsix}[1]{{\color{black}#1}}
\newcommand{\commentseven}[1]{{\color{black}#1}}
\newcommand{\ClarityText}[1]{{\color{black}#1}}

\usepackage[available,functional,reproduced]{usenixbadges}
\begin{document}


\title{\Large \bf Enabling Low-Cost Secure Computing on Untrusted
In-Memory Architectures 
}

\author{
{\rm Sahar Ghoflsaz Ghinani, Jingyao Zhang, Elaheh Sadredini}\\
University of California, Riverside\\ {\rm \{sghof001, jzhan502, elahehs\}@ucr.edu}
} 

\maketitle
\begin{abstract}
Modern computing systems are limited in performance by the memory bandwidth available to processors, a problem known as the memory wall. Processing-in-Memory (PIM) promises to substantially improve this problem by moving processing closer to the data, improving effective data bandwidth, and leading to superior performance on memory-intensive workloads. However, integrating PIM modules within a secure computing system raises an interesting challenge: unencrypted data has to move off-chip to the PIM, exposing the data to attackers and breaking assumptions on Trusted Computing Bases (TCBs). To tackle this challenge, this paper leverages multi-party computation (MPC) techniques, specifically arithmetic secret sharing and Yao’s garbled circuits, to outsource bandwidth-intensive computation securely to PIM.
\ClarityText{Additionally, we leverage precomputation optimization to prevent the CPU's portion of the MPC from becoming a bottleneck.} We evaluate our approach using the UPMEM PIM system over various applications such as Deep Learning Recommendation Model inference and Logistic Regression. Our evaluations demonstrate up to a \ClarityText{$14.66\times$} speedup compared to a secure CPU configuration while maintaining data confidentiality and integrity when outsourcing linear and/or nonlinear computation.

\end{abstract}

\section{Introduction}

\ClarityText{However, while processing power has grown exponentially following Moore's Law, memory performance has lagged, leading to the "memory wall"~\cite{Wulf1995-lm}—a critical bottleneck caused by the growing disparity between CPU and memory speeds, which limits overall system performance.} 

\commentseven{For data-centric applications, the cost of memory access is particularly significant, both in terms of latency and energy consumption~\cite{https://doi.org/10.48550/arxiv.2105.03814}. Boroumand et al.~\cite{consume} demonstrated that in Google consumer workloads, such as machine learning frameworks, 62.7\% of the total energy consumption stems from data movement between the CPU and memory. They further showed that Processing-in-Memory (PIM) can reduce total system energy consumption and execution time by an average of 55\% and 54\%, respectively, highlighting PIM’s potential as a solution to the memory wall challenge.}

\commentseven{PIM technology~\cite{tesseract, rowclone, MUTLU201928, mutlu2021,8686556,Chi2016-vn,Ahn2015-bf,zhang2023bp,zhang2022inhale,lenjani2020fulcrum,sadredini2021sunder} reduces memory latency by performing computations closer to or within the memory itself, leveraging internal data bandwidth. By minimizing data movement, PIM can significantly alleviate the limitations imposed by the memory wall. However, as promising as PIM technology appears, it comes with its own challenges, particularly in securely handling sensitive data~\cite{10.1145/3386263.3411365}.}

\begin{figure}[htbp]
\vspace{-1em}
\centerline{\includegraphics[width=3.1in, trim={0cm 0.5cm 0cm 0cm}]{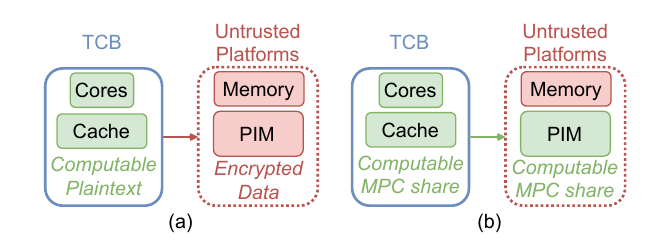}}
\caption{(a) TEE-based system. Only on-chip modules (e.g., cores, cache) are within the TCB. The PIM module cannot operate on encrypted data.
(b) TEE-based system with MPC. 
Private data can be split into two on-chip and off-chip shares. The PIM module can operate on computable MPC share.}
\label{fig:tee-mpc}
\end{figure}

\commentseven{Since PIM modules are typically off-chip and connected via buses such as memory or PCIe~\cite{mutlu2021,UPMEM,Ghose2018EnablingTA,sadredini2019eap,zhang2022sealer}, they are inherently outside the trusted computing base (TCB) established by widely deployed Trusted Execution Environments (TEEs)\cite{Kaplan2016-uv,Ngabonziza2016-zs,Costan2016-zb,Cheng2023-gq}. TEEs prohibit unencrypted data from appearing off-chip unless through a high-overhead attestation process\cite{Harriman2024TEEs}. The presence of plaintext in off-chip locations such as memory, memory buses, and peripheral interfaces makes it vulnerable to physical attacks (e.g., cold boot attacks~\cite{Halderman2009-vw}) and software attacks (e.g., unauthorized memory access~\cite{Lhee2003-qu,Tang2015-gn}). As shown in Fig.~\ref{fig:tee-mpc}(a), only on-chip modules, such as cores and caches, are included within the TCB. These modules can operate on plaintext, while off-chip modules like PIM can only handle encrypted data. Consequently, PIM hardware cannot natively compute on encrypted data, presenting a significant obstacle to its adoption in security-critical applications.}

\ClarityText{Although homomorphic encryption~\cite{hencryption,hetwo,hefour,hethree,homoup,gupta2023evaluating} allows computation directly on ciphertext, its significant computational complexity and data bandwidth requirements make it impractical for real-time applications, as it operates orders of magnitude slower than plaintext computation~\cite{Acar2018-se,Naehrig2011-ga}. An alternative is incorporating PIM into the TCB~\cite{Graviton,hix,HETEE}, but this would require substantial redesigns of the PIM architecture, such as adding a root of trust and secure boot mechanisms, as well as addressing supply chain security issues.}

\begin{figure}[htp]
\centerline{\includegraphics[width=2.7in, trim={0cm 0.8cm 0cm 0cm}]{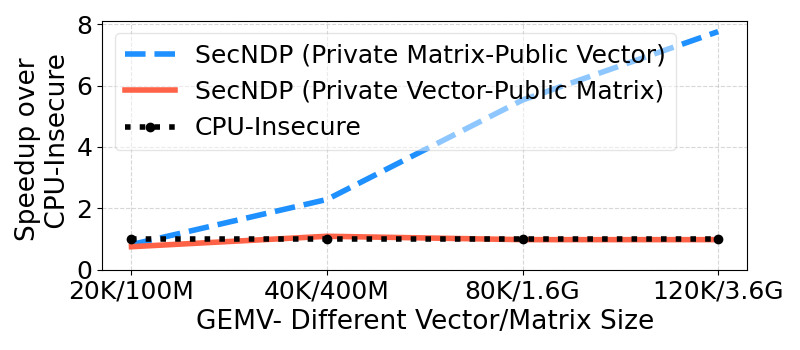}}
\caption{\ClarityText{Speedup of the SecNDP security scheme using UPMEM over an insecure CPU for the GEMV kernel with different input sizes.
The SecNDP does not perform well when the amount of public data is significantly large.}}
\label{secev}
\end{figure}

\commentseven{Xiong et al.\cite{9773244} proposed an MPC-based scheme, SecNDP, to securely offload linear computations to PIM. However, SecNDP has several limitations: (1) it does not support nonlinear computations, restricting its applicability; (2) its evaluation relies on simulations rather than real hardware, leaving its practicality in real-world scenarios unverified; and (3) our analysis reveals that SecNDP suffers from performance bottlenecks when the volume of public data significantly exceeds that of private data, as shown in Fig. \ref{secev}.}

\ClarityText{We analyzed the performance of SecNDP compared to an insecure CPU for GEMV with varying input sizes, as shown in Fig.~\ref{secev}. In GEMV, either the matrix or the vector can serve as private data, depending on the application. The blue line in Fig. \ref{secev} depicts the speedup when the matrix is private and the vector is public, while the red line shows the opposite case. The results demonstrate that SecNDP performs optimally when the public data size is smaller than the private data. However, as the public data size increases, CPU computation becomes a bottleneck due to the need to transfer large volumes of public data from memory to the CPU for processing, significantly constraining performance.}

\commentseven{To overcome these limitations, we propose a novel secure computation framework that leverages multi-party computation (MPC) to securely offload both linear and nonlinear computations to untrusted PIM hardware. As illustrated in Fig.\ref{fig:tee-mpc}(b), our method partitions plaintext into two encrypted shares: computationally intensive tasks are securely offloaded to the PIM, while lighter tasks are processed by the TEE. To address CPU bottlenecks, we employ precomputation techniques\cite{slalom} that shifts expensive tasks offline, significantly reducing runtime overhead. For secure data handling, we use counter-mode encryption~\cite{cmode1,cmode2} for parallelized encryption and Message Authentication Codes (MACs) to ensure data and computation integrity.}

\commentseven{Unlike SecNDP, our approach supports nonlinear computations and is validated on real-world PIM hardware (UPMEM)\cite{Devaux2019TheTP}, demonstrating its practicality and efficiency. Our framework is evaluated on diverse applications, including Deep Learning Recommendation Models (DLRM)\cite{Naumov2019-ep}, Multilayer Perceptrons (MLP)\cite{Pal1992-pn}, Linear Regression\cite{linreg}, and Logistic Regression~\cite{logreg}, showcasing its versatility and significant performance improvements.}

Our work makes the following contributions: 
\commentseven{
\begin{enumerate}[nosep]
    
    \item \textbf{Secure PIM Acceleration Framework:} We present a secure PIM acceleration framework leveraging MPC techniques, integrating arithmetic secret sharing for linear computations and Yao’s garbled circuits for nonlinear operations. By partitioning workloads into encrypted shares, our approach securely offloads computationally intensive tasks to untrusted PIM hardware. A switching mechanism between arithmetic secret sharing and Yao sharing is adopted to outsource nonlinear computation to PIM module securely.
    
    \item \textbf{Precomputation Optimization:} We utilize a precomputation technique to alleviate the CPU computational bottleneck in MPC by performing operations offline, reducing runtime overhead and enhancing the efficiency of secure data processing.

    \item \textbf{Empirical Evaluation on Real Hardware:} To the best of our knowledge, this is the first work to evaluate MPC-based secure computation on real-world PIM hardware (UPMEM). Our implementation and evaluation are publicly available\footnote{\href{https://zenodo.org/records/14736863}{https://zenodo.org/records/14736863} }, allowing other researchers to build upon our findings.

    \item \textbf{Comprehensive Workload Evaluation:} We extend our secure PIM framework to various real-world applications, including MLP inference, DLRM inference, Linear Regression (training and inference), and Logistic Regression (training and inference). These evaluations highlight our design's versatility in securely handling both linear and nonlinear computations.

    \item \textbf{Performance and Security Achievements :} Compared to a secure CPU implementation, our framework achieves speedups of $14.66\times$, $9.80\times$, $2.64\times$, and $5.85\times$ for MLP inference, DLRM inference, Linear Regression training, and Logistic Regression training, respectively. Our secure PIM design incurs only a minimal performance overhead (4\%) compared to an insecure PIM design, ensuring robust security without compromising efficiency.
\end{enumerate}}
\section{Threat Model} \label{sectionthreat}

\begin{figure}[htp]
\centerline{\includegraphics[width=2.7in]{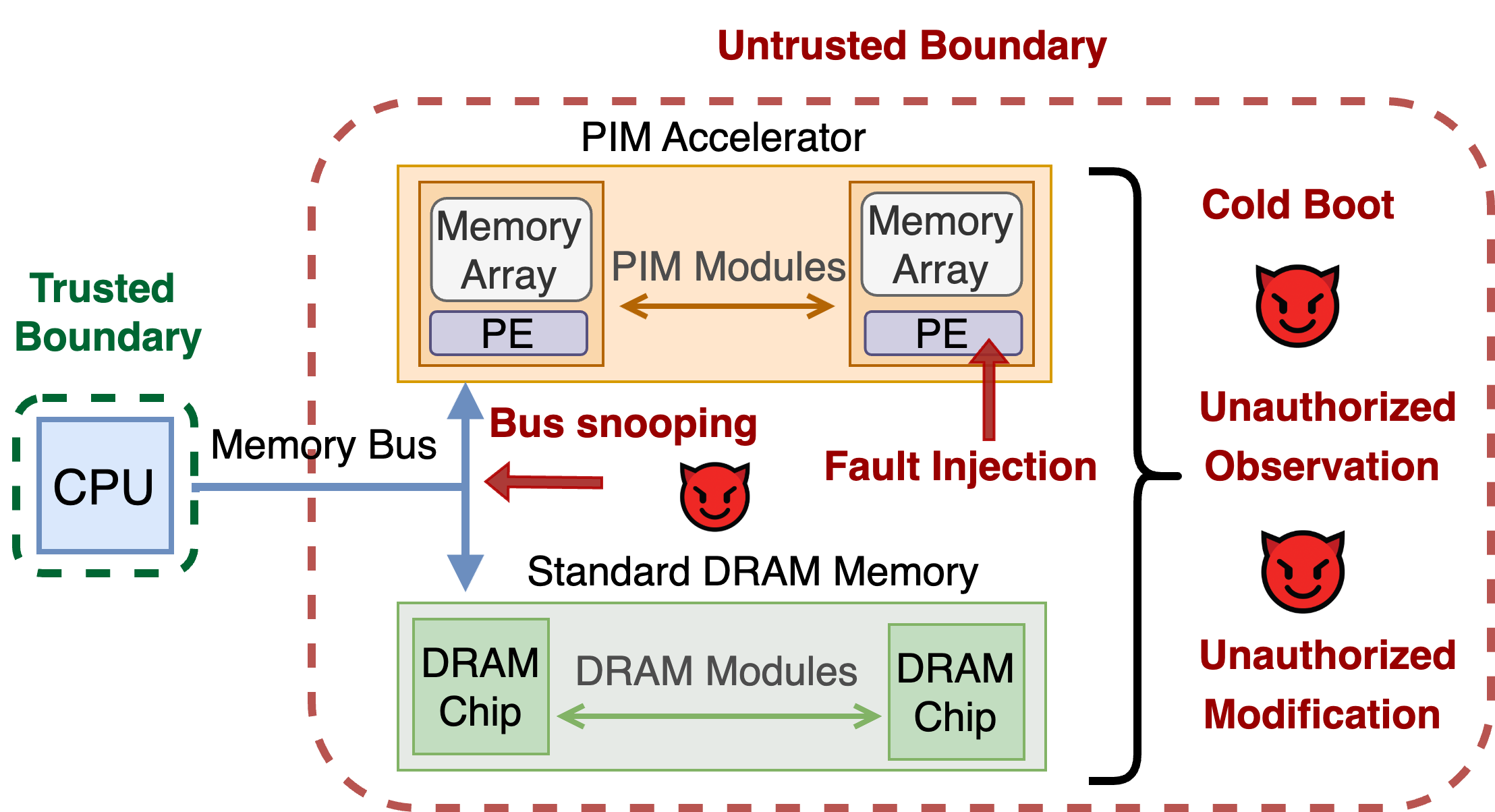}}
\vspace{-0.3cm}
\caption{Proposed threat model. TEE is the trusted party, and both standard and PIM-enabled memories are untrusted. An attacker can access or modify the data within these memories and the bus.}
\label{threat}
\end{figure}

Fig.~\ref{threat} presents our threat model, which aligns with standard TEE threat models commonly used in secure computing. 
Data within a TEE and the internal state of the processor are protected from unauthorized observation or modifications. Also, any malicious software present within the same processor cannot retrieve or modify private data.
However, all off-chip components are regarded as untrusted and vulnerable to attacks ~\cite{tee1,tee2}.

An attacker may attempt to access unencrypted user data stored in either standard or PIM-enabled memory or make unauthorized modifications. Additionally, attackers might eavesdrop on the bus to extract sensitive user information or tamper with the data. Furthermore, an attacker might launch a cold-boot attack~\cite{Halderman2009-vw} on the standard DRAM and the PIM-enabled DRAM to retrieve useful information. There could be vulnerabilities that enable Processing Elements in the untrusted PIM accelerator to inject faults into the computation. 

\ClarityText{As the computation involves only a single untrusted party, the possibility of collusion between parties is not relevant to our work.}
Additionally, side-channel attacks that exploit physical leakages, such as power analysis attacks~\cite{powtim}, timing attacks~\cite{powtim}, and electromagnetic attacks~\cite{elc}, are all excluded in our assumed threat model.

\vspace{-0.5cm}
\section{Background}\label{background}
\subsection{Processing-in-Memory (PIM)} \label{pim-back}
Processing-in-Memory, or \ClarityText{PIM~\cite{tesseract, rowclone, MUTLU201928, mutlu2021,8686556,Chi2016-vn,Ahn2015-bf}}, is an architectural paradigm that seeks to address the growing gap between the speed of processors and memory latency, referred to as the "memory wall"~\cite{Wulf1995-lm}. This strategy reduces or eliminates the need for data movement between the memory and the processing unit, leading to significant reductions in latency and energy consumption. PIM achieves this by incorporating computational capabilities into the memory unit, enabling computations to be performed right where the data resides.

PIM can be broadly classified into two categories based on where the computation takes place: Near Data Processing (NDP)~\cite{dlrmimp2,9773244} and In-Memory Computing (IMC)~\cite{8686556,Ahn2015-bf,Chi2016-vn}. The NDP approach, such as the one employed by UPMEM~\cite{Devaux2019TheTP}, places the computation units near the memory to minimize data movement. On the other hand, the IMC approach integrates the computational units directly into the memory chips, enabling computations to be performed in parallel across multiple memory units. 
Although IMC theoretically can further reduce data movement compared to NDP, it has not been widely adopted in commercial applications~\cite{Shanbhag2022-gl}. This is mainly due to the challenges associated with ensuring robustness during analog computation and integrating the system into the current computer system.

\subsection{UPMEM PIM Architecture}\label{upmem}
\begin{figure}[htp]
\centerline{\includegraphics[width=2.6in]{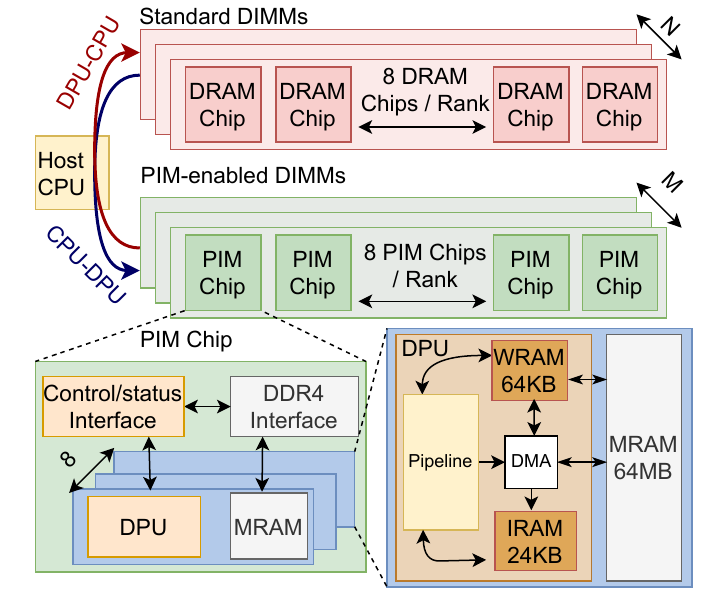}}
\vspace{-0.3cm}
\caption{Overview of the UPMEM Architecture. 
}
\label{upmemarch}
\end{figure}

UPMEM~\cite{Devaux2019TheTP, https://doi.org/10.48550/arxiv.2105.03814, 9651614, https://doi.org/10.48550/arxiv.2201.05072, 273853, 9908126,gómezluna2023experimental} is a promising example of the NDP approach. 
It combines the functionalities of standard DDR4-2400 DIMMs with specialized Processing Elements referred to as DRAM Processing Units (DPUs). The UPMEM DRAM chip is designed with two ranks, and each rank is equipped with eight PIM chips. As shown in Fig.~\ref{upmemarch}, each individual UPMEM PIM chip consists of eight banks, with each bank integrating a 64 MB Main RAM (\textit{MRAM}) and a DPU.
Furthermore, the inherent design of each DPU involves the utilization of 24 hardware threads known as tasklets, which operate in parallel within the DPU. These tasklets operate on the Single Instruction Multiple Data (SIMD) principles, executing the same task concurrently on various data sets.

The DPU chip incorporates two essential on-chip memory elements, namely Instruction RAM (\textit{IRAM}) and Working RAM (\textit{WRAM}), with the specific purpose of reducing the access time to the Main RAM (\textit{MRAM}) for the DPUs. 
The IRAM, a 24 KB memory, can store DPUs' encoded instructions, and the 64 KB WRAM, is accessible by DPUs using data store and load instructions. Additionally, DMA is utilized to enable communication between the MRAM and both the WRAM and IRAM.
The architecture, depicted in Fig.~\ref{upmemarch}, comprises two distinct types of DIMMs. Firstly, the standard DRAM functions as the primary memory for the host system. Secondly, there are PIM-enabled DRAMs, which can accelerate the computations in the memory. In this configuration, data is initially stored in the conventional main memory, and to perform computations on this data, it must be transferred to the PIM (i.e., CPU-DPU Communication).
Once the computation process is completed, the outcomes must be retrieved from all the DPUs, which entails DPU-CPU Communication. Given that these two memory types exhibit distinct data layouts~\cite{https://doi.org/10.48550/arxiv.2201.05072}, data transfer is performed through the CPU host \ClarityText{to adjust the layout}.

To minimize communication, public data can be transferred to the PIM before runtime, leaving only user-specific or updated data to be transferred during runtime.

\subsection{Trusted Execution Environment Extension}
Trusted Execution Environment (TEE) is a secure area inside a main processor. It guarantees code and data loaded inside are protected with respect to confidentiality and integrity. In the scope of TEE, memory encryption technologies play a vital role in ensuring data security. AES is a commonly used memory encryption technology~\cite{Kaplan2016-uv,Costan2016-zb, tiny-aes-c}.
By using AES, data that leaves the processor to be stored in the memory is encrypted, thus protecting the confidentiality of the information against potential attackers.
For the integrity of data - ensuring that it has not been altered or tampered with, a data structure known as a Merkle Tree is often employed~\cite{Kaplan2016-uv,Costan2016-zb}. By hashing the data and constructing a tree of hashes, the integrity of large sets of data can be verified efficiently and securely.

While memory encryption techniques within TEE can protect plaintext from being accessed by untrusted off-chip modules, it also means that these untrusted off-chip accelerators can only obtain encrypted data. As a result, they cannot accelerate applications effectively since computations based on encrypted data yield incorrect results.
One potential workaround is to extend the TEE's Trusted Computing Base (TCB) to include these accelerators~\cite{Graviton,hix,HETEE}. Through a series of protocols, these accelerators can be brought within the scope of the TCB, thus gaining access to the plaintext data for processing. However, this solution is not without its challenges.
Firstly, it would require the use of specialized accelerators designed with a Root of Trust for attestation, a feature lacking in most existing accelerators. Moreover, validating the integrity of these accelerators' supply chains to ensure the absence of embedded backdoors or other malicious modifications would also be necessary, a process that can be logistically and practically challenging. In essence, expanding the TCB of TEEs to include accelerators may not be feasible with current hardware, which further emphasizes the need for innovative solutions in secure, high-performance computation.

MPC~\cite{10.1007/978-3-642-32009-5_38,slalom,9773244,dark,snn,cnet} is a way to offload the computation to the off-chip components without the need to trust additional units. 

\subsection{Multi-Party Computation (MPC)} \label{back:mpc}
MPC is a branch of cryptography that allows parties to jointly compute a function over their inputs while keeping those inputs private. MPC provides an exceptional way to solve certain classes of problems where individual parties, each holding their own piece of private data, wish to collaborate on a shared computation without revealing their inputs to each other. It achieves this remarkable feat by dividing the sensitive data into multiple encrypted shares, each of which is processed independently.

The charm of MPC lies in its ability to execute computations on encrypted data, thereby ensuring the confidentiality of the data throughout the process. This feature makes MPC a potential solution for securely leveraging accelerators without the need to extend the TCB, since the computations on the accelerator side are still performed on encrypted data.

In a typical MPC scenario involving a TEE and accelerators, the sensitive data would be divided into two or more encrypted shares~\cite{slalom,9773244, 10.1007/978-3-642-32009-5_38,dark,10.1145/3243734.3243760}. One share would be processed on the TEE, and the rest on the accelerators. By \ClarityText{carefully} assigning computational tasks according to their complexity and volume - lightweight, low-volume tasks to the TEE and heavier, high-volume tasks to the accelerator - an efficient balance of the workload can be achieved.

\begin{figure}[htp]
\centerline{\includegraphics[width=2.7in]{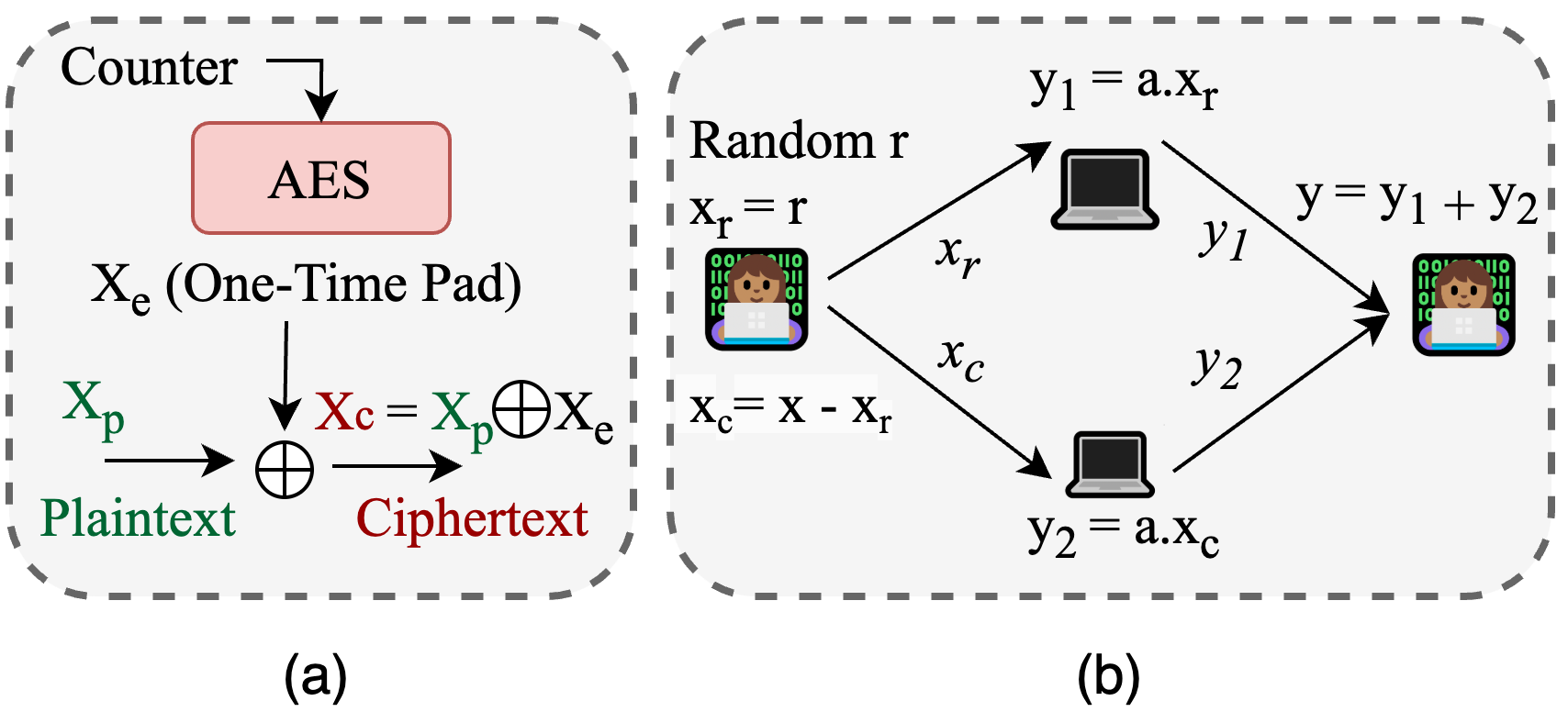}}
\vspace{-0.3cm}
\caption{(a) Counter-mode Encryption: Generated One-Time Pads are XORed with the plaintext to produce the ciphertext.
(b) Arithmetic Secret Sharing: The trusted party distributes the computation among multiple untrusted parties. Partial results are retrieved and aggregated to achieve the final result.
}
\label{cm}
\end{figure}

\textbf{Arithmetic secret sharing}~\cite{10.1007/978-3-642-32009-5_38,10.1145/3243734.3243760} is a type of Secure MPC where data is partitioned into multiple portions and arithmetic computations are performed separately by different parties over different partitions. Subsequently, the partial results obtained from each party are combined to derive the final result. An illustrative example of this process is presented in Fig.~\ref{cm}(b), where x is equal to $x_r + x_c$. The computation for $x_r$ is executed by party 1, while party 2 handles the computation for $x_c$. Finally, the partial results are aggregated to obtain the final result.
Each party is unaware of the actual data or the portion
held by other parties. Thus, the computation can be outsourced
while preserving privacy.

\textbf{Counter-mode encryption:} To employ counter-mode encryption~\cite{cmode1,cmode2}, as depicted in Fig.~\ref{cm}(a), the plaintext is not encrypted directly. Instead, a counter is encrypted to produce One-Time Pads (OTPs), a.k.a encrypted counter. A block of the plaintext is then XORed with the generated OTPs to produce a block of ciphertext. The decryption process simply requires an XOR operation. The encrypted counter is typically produced using a block cipher, such as AES. The CTR mode of AES is suitable for this purpose. One significant advantage of this counter-mode encryption is that it allows for parallel encryption of different blocks, due to the independence of the counters. This makes it an efficient method for symmetric encryption. Additionally, parts of the encryption and decryption processes can be precomputed. 

Combining arithmetic secret sharing with counter-mode encryption provides an effective and robust method for ensuring data confidentiality.

\textbf{Yao's garbled circuit:} 
Garbled Circuit (GC), as proposed by Yao~\cite{yao}, is a cryptographic protocol for secure two-party computation, typically employed when neither party trusts the other. In this scheme, one party constructs the GC and both parties then provide their encoded inputs to the circuit. Subsequently, one party can obtain the output from the GC. This output can also be shared with the other party.

Yao’s GC is computationally intensive due to the use of cryptographic primitives like AES. However, some optimizations like using free XOR gates~\cite{freexor} can help us to reduce or eliminate unnecessary cryptographic primitives, making GC more efficient for simple computations~\cite{yaosim,yaosim2}.

\subsection{Message Authentication Codes (MAC)}
Message Authentication Codes (MACs) are generated for each discrete block of data to detect potential unauthorized modifications. These algorithmically-produced tags are subsequently stored in memory alongside the corresponding data. Upon data retrieval, a new set of MACs is generated to be compared with the pre-existing, retrieved tags. Any discrepancies between these two sets of MACs would indicate a potential instance of unauthorized modification. In the context of linear functions, these tags can be generated via linear checksums or hash functions, which serve to map larger data blocks onto smaller data blocks. The likelihood of identical tags being generated for two distinct messages under this system is statistically negligible \cite{9773244}.

\vspace{-0.2cm}
\section{Related Work}

Numerous cryptographic techniques have been proposed to preserve the confidentiality of sensitive data and ensure computation integrity when outsourcing computations to untrusted parties.
While Homomorphic Encryption (HE)~\cite{hencryption,hetwo,hefour,hethree,homoup} is regarded as a promising solution for achieving high theoretical privacy, it is computationally intensive. Studies~\cite{survay} indicate that HE algorithms can introduce an overhead as high as two orders of magnitude in the best circumstances, making it impractical for secure computation using PIM. 

Gupta et al.~\cite{gupta2023evaluating} implemented and evaluated homomorphic operations on the UPMEM architecture.
While they significantly improved CPU and GPU implementations of homomorphic operations, their approach is still slower than MPC or TEE implementations. A detailed comparison is provided in section \ref{sec:eval}.
Differential Privacy (DP)~\cite{df1,df2} represents an alternative approach that safeguards data privacy by introducing controlled noise during computation, which in turn leads to a reduction in utility ~\cite{survay}. Increasing the level of privacy may decrease data accuracy and usefulness.

TEE is a widespread technique deployed in the cloud to provide data confidentiality and integrity. Acceleration can be achieved by extending TEE to untrusted off-chip modules ~\cite{Graviton,hix,HETEE}. 
Graviton~\cite{Graviton} and HIX~\cite{hix} propose GPU enclaves for secure GPU acceleration. However, they require hardware modifications, including hardware support for GPU enclaves and memory-mapped IO access protection.
HETEE~\cite{HETEE} uses an FPGA-based access control module to support TEE for PCIe-attached accelerators. However, HETEE does not protect the internal PCIe bus from physical attack and cannot support accelerators attached to the memory bus (e.g., NDP).

Additionally, MPC~\cite{10.1007/978-3-642-32009-5_38,slalom,9773244,dark,snn,cnet} is another noteworthy solution that necessitates trust in the central component while avoiding collusion. By combining MPC and TEE, secure acceleration can be achieved with minimal overhead, effectively balancing the need for data privacy and computational efficiency.

\ClarityText{Slalom~\cite{slalom} uses MPC to offload the linear computation to a co-located but untrusted GPU. While they are using precomputation to perform the computation on the TEE side, their scheme involves a significant communication overhead between memory, CPU, and GPU. Also, their scheme is only implemented for neural network inference.}

\ClarityText{SecNDP~\cite{9773244} proposes an MPC scheme to securely offload linear computation to an untrusted NDP. However, TEE computation can become a bottleneck when dealing with applications where public data is significantly larger than private data. In addition, their evaluation is based on simulation, and they do not support nonlinear computation.}

\ClarityText{\textit{To address these challenges, we propose a low-overhead security scheme to enable secure and fast computation of linear and nonlinear operations for off-chip PIM modules.}}
\section{\ClarityText{Proposed Secure PIM Computation }}

\subsection{Overview}

\ClarityText{This section provides a comprehensive overview of the encryption and verification mechanisms integral to our proposed framework for secure data processing on untrusted PIM architectures. The design ensures that sensitive data remains confidential while enabling secure computation offloading to PIM modules. Additionally, our verification scheme guarantees the integrity of data and computations outside the TEE.}

\vspace{-0.2cm}
\subsection{Preliminaries and Notations}

\ClarityText{To illustrate our proposed encryption and computation scheme, we use the MLP as an example, though the methodology is adaptable to other applications. The main computational bottleneck in many data-intensive tasks, including MLP, lies in linear operations such as GEMV, which we securely offload to PIM for acceleration.

Our scheme handles two types of data: private data, which must remain confidential, and public data, which does not require secrecy. Table~\ref{tab:notations} summarizes the key notations used in our framework.}

\ClarityText{Our encryption scheme utilizes One-Time Pads (OTPs) $(R_i)$ to securely mask plaintext $(P_i)$ during processing. Each OTP is generated using a unique key and a version number, ensuring that the masked data remains indistinguishable and secure, even if intercepted. This approach provides strong confidentiality with an extremely low risk of collisions~\cite{9773244}.}

\begin{table}[htp]
\vspace{-2mm}
	\centering
	\caption{List of Notations}
	\scalebox{0.85}{
	\begin{tabular}{>{\centering\arraybackslash}p{0.5in}>{\centering\arraybackslash}p{2.9in}}

		\toprule
  Notations&Explanation\\
		\midrule
  
    \textcolor{black}{$P_i$}& \textcolor{black}{$i^{th}$ element of the plaintext vector of size \textit{n}}\vspace{1mm}\\
    \textcolor{black}{$C_i$}& \textcolor{black}{$i^{th}$ element of the ciphertext vector of size \textit{n}}\vspace{1mm}\\
    \textcolor{black}{$R_i$}& \textcolor{black}{$i^{th}$ element of the generated OTP vector of size \textit{n}}\vspace{1mm}\\
    \textcolor{black}{$W_{j,i}$} & \textcolor{black}{Element at (j,i) of the weight matrix $(m \times n)$} \vspace{1mm}\\
     $resPIM$ & Retrieved partial results from PIM with size \textit{m}\vspace{1mm}\\
     $resCPU$ & CPU's partial results of size \textit{m}\vspace{1mm}\\
     $res$ & Final result ($resCPU + resPIM$) of size \textit{m}\vspace{1mm}\\
		\bottomrule
	\end{tabular}}
	\label{tab:notations}
    \vspace{-6mm}
\end{table}

\subsection{\ClarityText{Encryption Scheme}} \label{introEncrypt}

\ClarityText{Our encryption scheme employs arithmetic secret sharing~\cite{10.1007/978-3-642-32009-5_38} and counter-mode encryption~\cite{1431542, 10.5555/956417.956575, 10.1145/1152154.1152170} to securely offload linear computations to the PIM~\cite{9773244}. For nonlinear computations, which arithmetic secret sharing cannot handle, we utilize Yao's garbled circuits~\cite{yao1986generate, 10.1145/3243734.3243760, secureml}, ensuring secure processing for a wide range of workloads.}

\ClarityText{As explained in section~\ref{back:mpc}, before runtime, the plaintext is masked with OTPs ($R_i$) to generate the ciphertext ($C_i$), which is stored in memory for use during execution. In our scheme, PIM-enabled memory handles computation on the ciphertext ($C_i$) to reduce data movement between the TEE-enabled CPU and PIM. The TEE is responsible for performing computations over the OTPs ($R_i$), as $R_is$ can be generated on the fly with little or no memory access. The final result is obtained by combining partial results from PIM and TEE.}

\vspace{-0.2cm}
\vspace{0.1cm}
\subsubsection{Secure outsourcing of linear functions}\label{LinearOutsourcing}
Computation over OTPs can be performed either at runtime in parallel with the PIM computations, i.e., the runtime method, or precomputed offline, i.e., the precomputation method.

\begin{figure}[htp]
\centering
\includegraphics[width=2.6in, trim={0.5cm 0.5cm 0.5cm 0.5cm}]{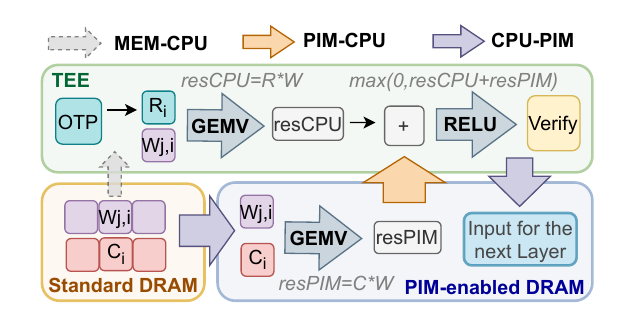}
 \caption{\ClarityText{The computation flow of the runtime approach begins with the initiation of a GEMV kernel by the PIM system on the ciphertext. Concurrently, on the CPU side, a set of OTPs is generated, and GEMV computation is performed on them. Finally, the partial results are merged within TEE.} }
\label{randp}
\end{figure}

\begin{figure}[htp]
\centerline{\includegraphics[width=2.81in]{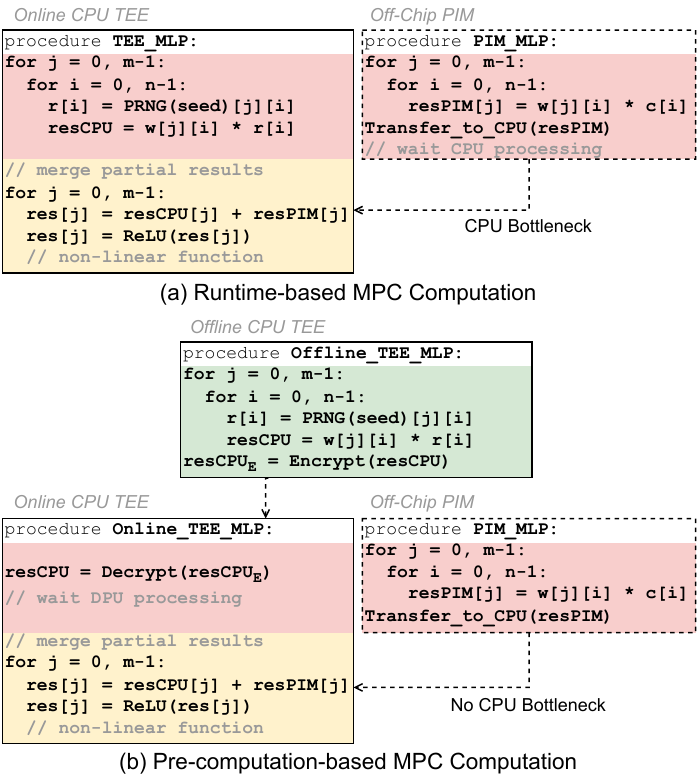}}
\vspace{-0.3cm}
\caption{\ClarityText{(a) In runtime-based MPC, the CPU generates OTPs and performs GEMV computations, which can create bottlenecks on the CPU side. (b) In precomputation-based MPC, these computationally intensive tasks are handled offline, removing them from the critical path and resulting in lightweight runtime CPU computation.
}}
\label{fig:precompute}
\vspace{-0.3cm}
\end{figure}

\ClarityText{\textbf{Runtime Scheme:}} Fig.~\ref{randp} shows the overview of computation using the runtime method. 
\ClarityText{Once the execution is started, the ciphertext ($C_i$), which is stored in the standard memory, is sent to the PIM. Meanwhile, OTPs are generated on the TEE-enabled CPU. Then, kernel computations are performed in parallel across the TEE and the PIM.
It is essential to ensure that the CPU does not become a performance bottleneck; otherwise, the overall computational benefits of using the PIM accelerator would be negated.
The CPU avoids becoming a bottleneck by generating OTPs on the fly, eliminating the need for additional memory accesses. This streamlined OTP generation enables the PIM to perform its operations at full capacity without waiting for CPU side computation.}
Once the computation on the PIM side is complete, the CPU retrieves partial results ($resPIM$). The final result is obtained by applying the ReLU function to the sum of CPU results and PIM results ($resCPU + resPIM$). \ClarityText{Fig.~\ref{fig:precompute}(a) presents the pseudo-code for runtime-based implementation of MLP.}

\ClarityText{The runtime method may be less effective in applications where the volume of public data significantly exceeds that of private data. In this case, while the runtime generation of OTPs obviates the need to transfer $C_i$ from memory to the CPU, a substantial amount of public data must be transferred, which causes the CPU computation to become the bottleneck.
We employ the precomputation scheme (explained next) to mitigate this issue.}

\commentfour{\textbf{Precomputation Scheme:}
The precomputation scheme can mitigate the issue associated with the runtime scheme by precomputing the kernel computation over OTPs on the TEE. This makes the runtime computation on the CPU lightweight and prevents it from becoming a bottleneck. The reason behind this is that OTPs are encrypted counters and can be generated independently of the private data, allowing us to precompute the kernel over them.}

\begin{figure}[htp]
\centerline{\includegraphics[width=2.7in, trim={0.5cm 0.3cm 0.5cm 0.5cm}]{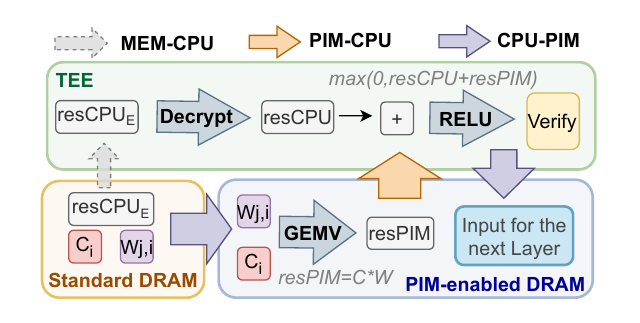}}
\vspace{-0.3cm}
\caption{ \ClarityText{Precomputed partial results ($resCPU$) are securely stored in memory. During runtime, the PIM accelerator performs computations over the ciphertext ($C_i$). However, the computation on the CPU is limited to decrypting the precomputed results ($resCPU$).
}}
\label{fig:precompute1}
\end{figure}

\commentfour{As shown in Fig.~\ref{fig:precompute}(b), the precomputation scheme consists of two phases: the offline and online phases. In the offline phase, the computation is performed on the OTPs, and the encrypted results are stored in memory. In the online phase, as presented in Fig.~\ref{fig:precompute1}, the computation on the TEE is as lightweight as decrypting the encrypted results ($resCPU$). The computation on the PIM remains the same as in the runtime scheme. Once the computation on the PIM is complete, the TEE merges the precomputed results ($resCPU$) with the partial results retrieved from the PIM ($resPIM$).
The precomputation method can be applied to applications such as neural network inference, where the weights do not change dynamically.}

\vspace{-0.4cm}
\subsubsection{\commenttwo{Secure outsourcing of non-linear functions}}\label{NLOutsourcing}

\begin{figure}[htp]
\centerline{\includegraphics[width=2.8in, trim={0cm 0.4cm 0cm 0.4cm}]{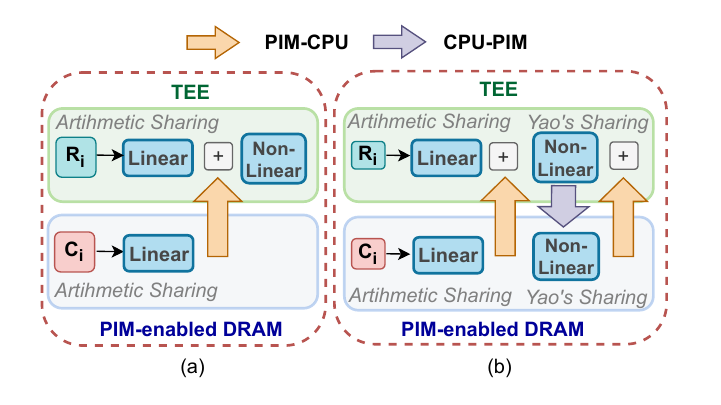}}
\vspace{-0.3cm}
\caption{(a) Securely outsourcing the linear computation to the PIM module.
(b) Securely outsourcing both linear and nonlinear functions to the PIM. The main difference between these two is that we need to switch to Yao sharing to outsource nonlinear functions securely.}
\label{fig:yaos}
\vspace{-0.3cm}
\end{figure}

\commenttwo{
To perform both linear and nonlinear computations securely on the PIM, we employ a switching technique that transitions from arithmetic secret sharing to Yao's Garbled Circuits (Yao sharing)~\cite{yao,yao1986generate,10.1145/3243734.3243760,secureml}. This dual approach enables our framework to handle a broader range of computations securely and efficiently.

When only linear computations are offloaded to the PIM (Fig.~\ref{fig:yaos}(a)), the system simply retrieves the intermediate results after computation. However, offloading both linear and nonlinear computations requires an additional step. As shown in Fig.~\ref{fig:yaos}(b), the TEE generates Yao's circuit based on the desired nonlinear function and shares it with the PIM. This process ensures that the nonlinear computation is securely handled on the untrusted PIM hardware.

To optimize the performance of nonlinear computations using Yao's circuits, it is advantageous to convert complex functions into more GC-friendly equivalents. Yao's circuits are inherently efficient for Boolean logic, and transforming nonlinear functions into circuit-compatible formats minimizes overhead. For example, the sigmoid function is computationally expensive to implement directly using Yao's Garbled Circuits. To address this, Mohassel et al.~\cite{secureml} proposed a GC-friendly approximation of sigmoid for logistic regression, defined as:

\setlength{\abovedisplayskip}{3pt}
\[
f(x) = 
\begin{cases} 
0, & \text{if } x < -\frac{1}{2}, \\
x + \frac{1}{2}, & \text{if } -\frac{1}{2} \leq x \leq \frac{1}{2}, \\
1, & \text{if } x > \frac{1}{2}.
\end{cases}
\]
\setlength{\belowdisplayskip}{3pt}

This function can be represented in a Boolean circuit-friendly format as:
\[
f(u) = (\neg b_2) + (b_2 \land (\neg b_1)) u,
\]
where 
\[
b_1 = 
\begin{cases} 
0, & \text{if } u + \frac{1}{2} \geq 0, \\
1, & \text{otherwise,}
\end{cases}
\]
and
\[
b_2 = 
\begin{cases} 
0, & \text{if } u - \frac{1}{2} \geq 0, \\
1, & \text{otherwise.}
\end{cases}
\]
The TEE computes \( b_1 \) and \( b_2 \), constructs the corresponding Garbled Circuit, and securely shares it with the PIM.

Once the Yao's circuit is constructed and shared, both the TEE and PIM feed their respective inputs into the circuit to compute the final result. Since the TEE is a trusted party, it processes plaintext data without encryption, reducing overhead compared to traditional Yao's Garbled Circuit implementations.

While this switching scheme introduces communication overhead due to the construction and sharing of Yao's circuits, the resulting acceleration in computation justifies the trade-off. By leveraging this dual approach—arithmetic secret sharing for linear tasks and Yao's Garbled Circuits for nonlinear computations—we enable the PIM to securely and efficiently handle diverse workloads, ensuring robust performance across a wide range of applications.
}

\vspace{-0.7cm}
\subsection{\ClarityText{Verification Scheme}}

Given that off-chip accelerators are untrusted, it is crucial to ensure computational integrity and detect any unauthorized modifications. 
We employ MACs to address this concern. To generate a MAC for a block of data, a linear checksum can be utilized. 
Specifically, we employ Linear Modular Hashing~\cite{cmode2,mac2,9773244} to generate tags for data blocks. For the GEMV kernel, the linear checksum can be formulated as below~\cite{9773244}:
\setlength{\abovedisplayskip}{2pt}
\[Tag_j = (\sum_{i=0}^{m-1} P_{i,j} \times s^{m-i} )\; mod \; q\]
\setlength{\belowdisplayskip}{2pt}
Verification tags \textit{($Tag_j$)} are generated for each column of the matrix. \textit{q} is a big prime number and \textit{s} is an encrypted value using a key, a version number \textit{v}, and the address of P. 

\begin{figure}[htp]
\centerline{\includegraphics[width=2.8in, trim={0.6cm 0.1cm 0.6cm 0.5cm}]
{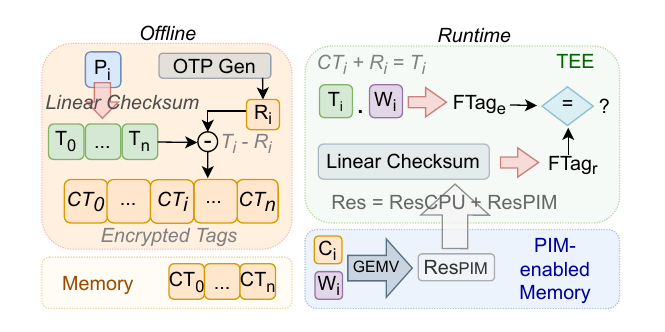}}
\vspace{-0.3cm}
\caption{In the offline phase, tags are generated and stored in memory. At runtime, the precomputed tags are decrypted, and the same kernel computation is performed on them ($FTag_{e}$). Meanwhile, new tags are generated based on the final result ($FTag_{r}$). If these tags are equal, the computation is verified.}
\label{verif}
\vspace{-0.4cm}
\end{figure}

As shown in Fig.~\ref{verif}, tags are generated and securely stored in the memory using MAC-then-encrypt strategy~\cite{mte}. At runtime, the encrypted tags are decrypted, and the same kernel computation is executed over the tags to generate $FTag_{e}$. 
Since the overhead of generating $FTag_{e}$ is negligible and can be parallelized with the PIM computation, we opt to generate them at the runtime on TEE.
Subsequently, we retrieve the computation results from PIM and merge them with the partial results from CPU ($res = resCPU + resPIM$). Applying Linear Modular Hashing to these results generates $FTag_{r}$. By comparing $FTag_{e}$ with $FTag_{r}$, we can verify the computation and ensure there are no unauthorized modifications.

The process requires multiple verification stages when offloading both linear and non-linear functions to PIM. For example, in the secure implementation of logistic regression, there are two steps of verification: 1) after the dot product and 2) after gradient descent. Depending on the application, the number of verification rounds may increase if we switch to arithmetic sharing multiple times.

 \vspace{-0.2cm}
\ClarityText{\subsection{Evaluation Baselines }}\label{otherworks}

\ClarityText{This section discusses the baseline approaches used for comparing against our proposed secure PIM framework.}

\begin{figure}[htp]
\centerline{\includegraphics[width=2.8in, trim={0.6cm 0.6cm 0.6cm 0.6cm}]{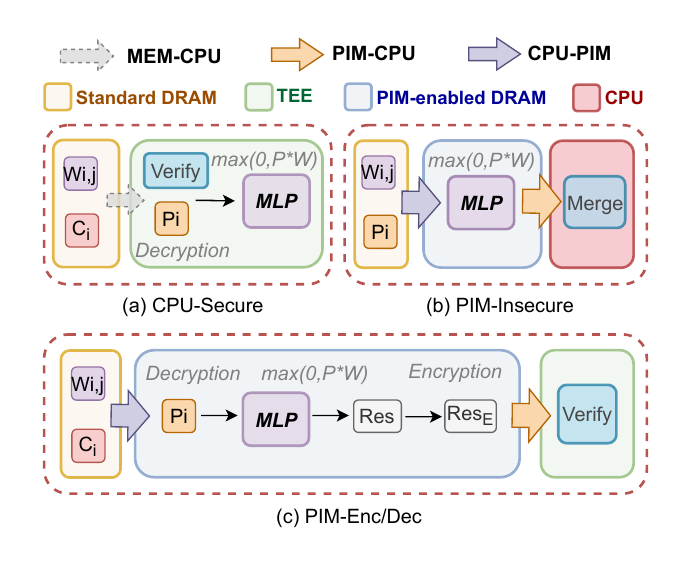}}
\vspace{-0.3cm}
\caption{(a) \ClarityText{CPU-Secure: TEE-based computation on the CPU.} The TEE decrypts the encrypted data, enabling it to carry out the required computations. (b) \ClarityText{PIM-Insecure:} The PIM accelerator directly performs computation on the plaintext.  (c) \ClarityText{PIM-Enc/Dec:} The PIM accelerator first decrypts the encrypted data to perform the computation.
}
\label{tee-flow}
\end{figure}

\ClarityText{As illustrated in Fig.~\ref{tee-flow}(a), the CPU-Secure baseline represents a system with a TEE-enabled CPU that performs computations on encrypted data stored in main memory. The TEE-enabled CPU decrypts the data, processes it, and ensures confidentiality and integrity. However, frequent data movement between memory and the CPU introduces significant overhead, which limits the overall system efficiency.}

\ClarityText{The PIM-Insecure baseline, shown in Fig.~\ref{tee-flow}(b), transfers plaintext data from main memory to the PIM for computation. While this approach leverages PIM-based computation to reduce data movement and achieve high performance, it completely lacks security mechanisms, exposing sensitive data to potential risks. Additionally, handling both linear and nonlinear computations in this setup increases communication overhead between the CPU and PIM.}

\ClarityText{In Fig.~\ref{tee-flow}(c), the PIM-Enc/Dec baseline represents a system where the PIM and CPU are assumed to share encryption keys for secure data exchange. The CPU encrypts data before transferring it to the PIM, where it is decrypted, processed, and re-encrypted before returning the results to the CPU. Although this ensures data confidentiality and integrity, the repeated encryption and decryption operations introduce substantial performance overhead, particularly for large datasets. This scheme is partially secure based on the TEE's threat model, as it assumes that the PIM is trusted.}

\ClarityText{Our proposed framework overcomes these limitations by securely offloading computations to untrusted PIM hardware without compromising efficiency, achieving a balance between robust security and high performance.}

Fig.~\ref{tee-flow}(c) provides an overview of our secure PIM baseline, assuming that both the CPU and the Processing Elements (PEs) within a PIM chip are trusted entities that share a secret key. This key is utilized to encrypt and decrypt data stored in the memory. 
The PIM loads the encrypted data onto the PEs and decrypts it to perform the computation.
Afterward, the PEs encrypt the results before sending them back to the CPU. Compared to our scheme, this approach requires additional encryption and decryption steps, which affects performance when handling large private datasets. Furthermore, it processes unencrypted data in the PIM, which contrasts with the TEEs' threat model.

\vspace{-4mm}
\commentthree{
\subsection{Security Argument}

Our security scheme is built upon existing cryptographic techniques introduced in SecNDP~\cite{9773244}, Slalom~\cite{slalom}, and SecureML~\cite{secureml} to ensure the privacy and integrity of data and computation while offloading them to the PIM. 

\textbf{Security Guarantees:} The following cryptographic techniques are used in our scheme:
\begin{itemize}[nosep]

\item \textbf{Privacy of Data:} Counter-mode encryption (CTR) is employed to ensure the privacy of sensitive data by masking it with OTPs. CTR mode uses a block cipher, such as AES, to generate encrypted counters, which serve as OTPs. These OTPs are generated using random numbers that are securely and honestly produced within the TEE. As a result, masking the plaintext with these OTPs makes the data indistinguishable from random values from the adversary's perspective. These security guarantees are based on the assumptions validated in SecNDP\cite{9773244}.

\item \textbf{Correctness of Linear Computation}: Arithmetic secret sharing is used to ensure that the linear computation is performed privately by splitting it into two unrecognizable parts between the PIM and TEE. Each share alone cannot reveal any information. The correctness of linear computation is based on the assumptions validated in \cite{9773244, slalom}.

\item \textbf{The Correctness of Non-linear Computation}: Yao's Garbled Circuit is employed to perform non-linear computations securely using the PIM and TEE. SecureML \cite{secureml} outlines the cryptographic guarantees and correctness guarantees for this technique.

\item \textbf{Verification and Integrity}: Message Authentication Codes (MACs) are used to detect any unauthorized modifications and to verify computations on the PIM side. We use a linear modular hashing technique to verify linear computations. SecNDP \cite{9773244} presents how this scheme verifies computation and ensures integrity.

\end{itemize}}

\vspace{-3mm}
\section{Evaluation Methodology}\label{sec:eval-method}
\vspace{-2mm}
\subsection{Workload}\label{logist}
This section describes our use cases and how the computation is distributed among different DPUs. We have implemented MLP inference, DLRM inference, Logistic Regression training, and Linear Regression (training and inference).

\textbf{Implementation of MLP:}
MLP~\cite{10.5555/646424.692266} is a class of neural networks with at least three layers. Typically, each layer consists of a GEMV kernel followed by the activation function (ReLU)~\cite{https://doi.org/10.48550/arxiv.2105.03814}.
In our baseline UPMEM implementation of MLP inference~\cite{https://doi.org/10.48550/arxiv.2105.03814, gomez2021benchmarking, mlpimplem}, each layer of the MLP involves a weight matrix and an input vector. We consider the vector inputs to be private and weights to be public.

\begin{figure}[htp]
\centerline{\includegraphics[width=3in, trim={0cm 0.5cm 0cm 0.5cm}]{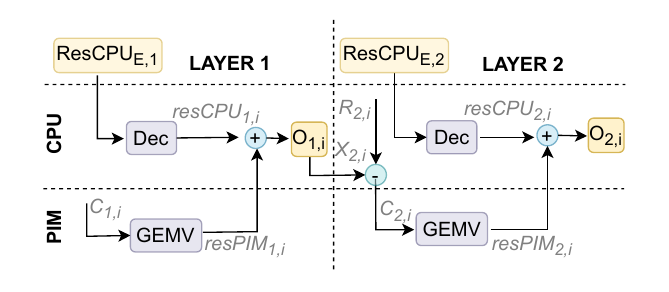}}
\vspace{-0.3cm}
\caption{An example of how precomputation can be used for MLP with multiple layers of computation. Since the output of each layer is unencrypted, it must be encrypted before being passed to the next layer.
}
\label{fig:new}
\vspace{-0.3cm}
\end{figure}

In this setup, to optimize the MLP computation, only the GEMV operation is offloaded to the UPMEM, while the CPU handles all other computations. The matrix is divided into equal row-wise partitions and distributed among the DPUs. Meanwhile, a copy of the vector is sent to each DPU.
In MLP inference, each layer's output serves as the next layer's input. As shown in Fig.~\ref{fig:new}, the input (\(X_{1,i}\)) is secret-shared between the CPU and PIM (\(C_{1,i} = X_{1,i} - R_{1,i}\)). Then, the PIM computes \(ResPIM_{1,i} = C_{1,i} \times W_{1,i}\), while \(ResCPU_{1,i}\) is precomputed offline in the CPU. The TEE merges these partial results to obtain the input for the next layer (\(X_{2,i} = ResCPU_{1,i} + ResPIM_{1,i}\)). The TEE encrypts \(X_{2,i}\) by subtracting \(R_{L+1}\) from it and secret shares it with the PIM. The overhead associated with this subtraction is negligible since \(R_{2,i}\) is generated on the fly and \(X_{2,i}\) is present in the cache.

The implemented secure MLP consists of ten layers. We experimented with various input and weight dimensions (Fig.~\ref{fig:mlp-comp}(a)), as well as different batch sizes (Fig.~\ref{fig:mlp-comp}(b)).
Our implementation was evaluated with three different input sizes (1000, 5000, and 10000) and three different weight sizes ($1000\times1000$, $5000\times5000$, and $10000\times10000$). A weight dimension of $1000\times1000$ means that each layer consists of 1000 neurons, and each neuron is connected to 1000 neurons in the next layer.

\textbf{Implementation of DLRM:}
DLRM~\cite{Naumov2019-ep}, is used for personalization and recommendation tasks, distinguishes itself from other deep learning models by handling both continuous and categorical features as input. Continuous features are processed through a dense MLP, while millions of embedding vectors represent categorical features.

The embedding table lookup operation involves a weighted sum, involving a sparse lookup followed by a reduction operation. This reduction operation corresponds to the Pooling Factor (PF), or the aggregation size. Given a batch size denoted by \textit{B}, a list of indices $I = \{id_0, id_1, ..., id_{B\times PF-1} \}$, and a list of weights \textit{A}, the output of the computation over the embedding table \textit{W} is given as below when $k=\{0, 1, ..., B-1\}$.
\[ Output_{k,i} = \sum_{j=k}^{(k*PF)-1} a_{(id)_j} \times w_{{(id)_j},i}\]

Our insecure PIM baselines, PIM-Rec~\cite{Zarif_2023,dlrmimp1,dlrmimp2}, is based on Meta's DLRM~\cite{Naumov2019-ep}, which primarily offload the embedding lookup table operation to UPMEM, while other operations are carried out on the CPU. This is due to the high degree of irregular memory accesses required by embedding lookup operation and the considerable amount of data involved. \cite{dlrmimp2} splits the columns of the embedding tables and offloads them to the DPUs. 

The user data within the embedding lookups is private, so it is essential to ensure that it is securely offloaded and processed in memory.
Similarly to our baselines, the model architecture can be configured to evaluate the design, and the input can be randomly generated~\cite{Zarif_2023}. For our test cases, the number of embedding lookup tables is 64, PF=32, and batch size is 128.

\textbf{Implementation of Logistic Regression:}
Logistic Regression~\cite{logreg}, is a method employed for binary classification. This statistical technique outputs the probability that an input belongs to one of the two classes.
In the training phase of logistic regression, numerous iterations are conducted to find the optimal parameters.
Each iteration of logistic regression training begins with a forward pass, which involves the dot product of the weights and inputs, followed by the activation function (e.g., sigmoid). Next, the loss function is calculated to evaluate the accuracy of our predictions, followed by the computation of gradient descent to update the model parameters.

In the implementation of logistic regression training, we adhere to the approach outlined in our UPMEM-Insecure baseline study~\cite{gómezluna2023experimental, pim-ml}. Initially, the input is distributed row-wise across PIM cores and tasklets. The CPU performs the final reduction step in each iteration and updates the model parameters. 

We implemented two different versions: one only executes linear functions on the PIM (UPMEM-Runtime-A), and the other performs both linear and nonlinear computation on the PIM cores (UPMEM-Runtime-A2Y).
\commenttwo{
In the second case, we must temporarily switch to Yao's sharing to perform the activation function. As discussed in section \ref{NLOutsourcing}, we utilize the GC-friendly activation function to be able to efficiently and securely offload non-linear functions to PIM.
}

Both schemes were evaluated over 1000 iterations with a learning rate of 0.001 and 16 features \textit{(n)}.

\textbf{Implementation of Linear Regression:}
Linear regression~\cite{linreg}, is a statistical technique used to model the optimal relationship between a dependent variable and one or more independent variables using a linear equation.
Like logistic regression, the training involves multiple iterations to find optimal parameters. Each iteration is identical to logistic regression except for the activation function. In the baseline implementation of linear regression~\cite{gómezluna2023experimental, pim-ml}, the input is distributed across PIM cores and tasklets. Partial results are then merged within DPUs and CPU before updating the weights. 
Since we execute both dot product and gradient descent on UPMEM, some additional communication between CPU and PIM, after dot products and before gradient descent, is necessary. 

Linear regression inference is simply a matrix-vector multiplication of the weights and inputs. The input matrix consists of several samples, each with different features. The weight vector has the same length as the number of features.

\vspace{-3mm}
\subsection{Hardware Settings}

We evaluated our method using UPMEM PIM hardware~\cite{UPMEM}, which is mainly the standard DDR4-2400~\cite{JEDEC} DIMM integrated with DPUs.
Each UPMEM DIMM has 128 DPUs, each with 64 MB of memory and a communication rate of 1 GB/s. Increasing the number of PIMs enhances parallel computing capabilities and throughput, but this is limited by memory capacity (server’s DIMM channels/slots)~\cite{UPMEM}. For our experiments, we used 20 PIM-enabled DIMMs, resulting in a total of 160 GB of MRAM, which is the maximum supported by a dual-socket Cascade Lake server (Intel Xeon Silver 4110 CPU~\cite{intel}). This configuration supports the parallel operation of 2,560 DPUs, all running at a clock frequency of 350 MHz. The CPU host can send data in parallel to the PIMs; thus, fewer PIMs result in reduced bandwidth.

UPMEM-Insecure represents the insecure version of our UPMEM baseline implementation for various applications, and UPMEM-Enc/Dec denotes a secure variant under the assumption that both the CPU and the DPUs are trusted (Described in section~\ref{otherworks}). In addition, for UPMEM-Enc/Dec, the DPUs can perform both encryption and decryption operations using AES at a rate of 5 MB/sec. 
Our proposed framework can be applied to any PIM-based architecture as it is independent of the specific PIM hardware specifications.

\commentone{CPU-Insecure and CPU-Secure are evaluated on the same system as UPMEM-Insecure and UPMEM-Secure. In the CPU-Secure version, TEE protection is provided using Intel's Trust Domain Extensions (TDX), a VM-based TEE that offers improved efficiency compared to hardware-based TEEs.}

\vspace{-3mm}
\section{Evaluation Result} \label{sec:eval}
\vspace{-0.2cm}

\subsection{Performance Analysis}
 In this section, we evaluate the latency and security analysis of our scheme using four memory-bound applications.

\begin{figure}[htp]
\centerline{\includegraphics[width=3.2in, trim={0cm 1cm 0cm 0cm}]{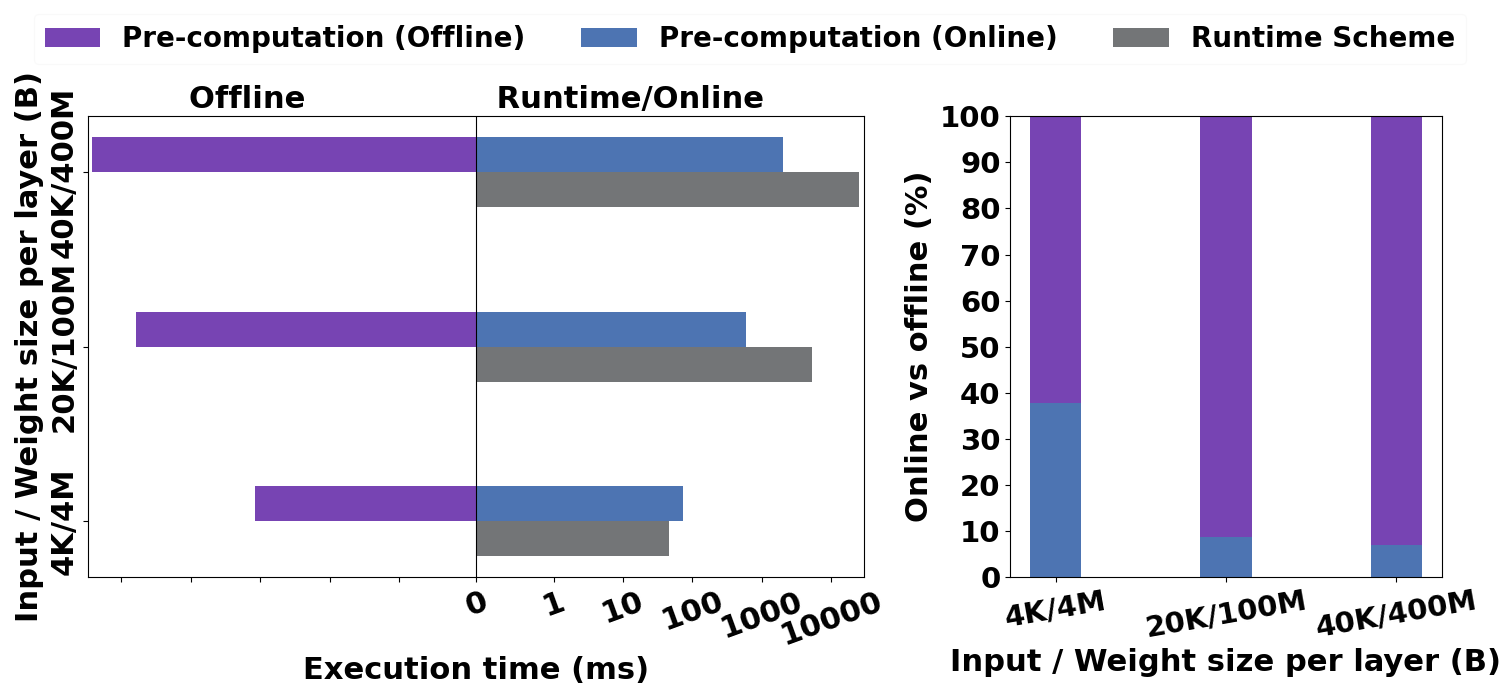}}
\caption{\commentfour{(Left) Comparing precomputation vs. runtime scheme for MLP. (Right) Online vs. offline proportion in precomputation scheme for MLP. 
}}
\label{pre-mlp}
\end{figure}

\textbf{Precomputation Overhead:}
\commentfour{
We evaluate the precomputation scheme on MLP. This scheme can be applied to applications where the weights remain static and do not change dynamically.
As described in Section \ref{LinearOutsourcing}, the precomputation scheme operates in two phases: offline and online. Figure \ref{pre-mlp} (left) illustrates the overhead of the offline phase in the precomputation scheme and compares the online phase of the precomputation scheme with the runtime scheme for MLP. By precomputing the computationally intensive portions offline, the execution time of MLP is significantly reduced, achieving up to a 92\% reduction compared to the runtime scheme. This demonstrates that with precomputation, the CPU portion of the computation no longer becomes a performance bottleneck.

Furthermore, increasing the input size for MLP primarily affects the offline phase (Fig \ref{pre-mlp} (right)). Since the offline phase consists of kernel operations over OTPs and the encryption of results, the associated overhead remains bounded.}

\begin{figure*}[!tp]
\centerline{\includegraphics[width=0.8\textwidth, trim={0cm 1cm 0cm 0cm}]{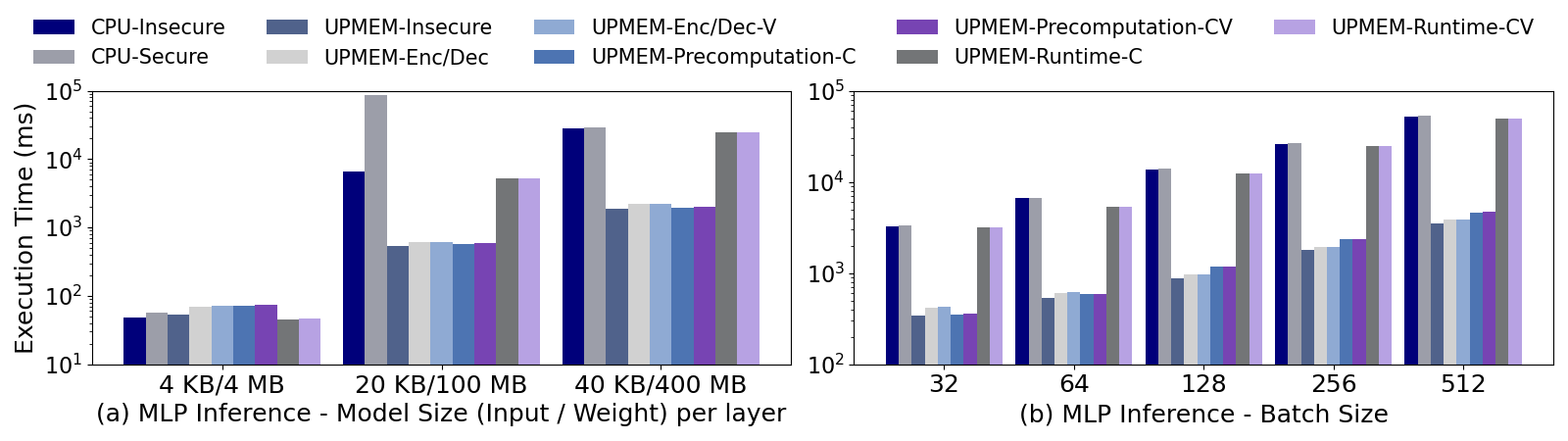}}
\caption{\commentone{{Comparing different MLP implementations: (a) with varying model sizes for a batch size of 64, where increasing the model size enables UPMEM-Precomputation to achieve greater speedups over CPU-Secure. (b) with varying batch sizes for a model size of 20 KB/100 MB per layer, where UPMEM-Precomputation consistently delivers the best performance.
}}}
\label{fig:mlp-comp}
\end{figure*}

\textbf{MLP performance analysis:}
Fig.~\ref{fig:mlp-comp}(a) compares the execution time of MLP under various operational modes: precomputation (UPMEM-Precomputation) and runtime (UPMEM-Runtime), against insecure CPU (CPU-Insecure), insecure UPMEM~\cite{https://doi.org/10.48550/arxiv.2105.03814} (UPMEM-Insecure), TEE-enabled CPU (CPU-Secure), and secure UPMEM with the assumption of having trusted DPUs (UPMEM-Enc/Dec). Utilizing 2496 DPUs and 16 tasklets, \commentone{our precomputation method can achieve a $14.28\times$ and a $14.66\times$ speedup over CPU-Insecure and CPU-Secure for the model with an input size of 40 KB/layer, respectively.} However, the runtime method has negligible speedup ($1.14 \times$) \ClarityText{compared to CPU-Insecure}, primarily due to the significant performance constraint imposed by the large volume of data transfer between the standard memory and the CPU for weights (public data).
\commentone{On average, the CPU-Secure implementation of MLP results in about $3\%$ overhead compared to CPU-Insecure.}

We incorporated batch processing to the baseline implementation of MLP~\cite{https://doi.org/10.48550/arxiv.2105.03814} (UPMEM-Insecure), leading to a $14.87\times$ speedup compared to CPU-Insecure for an input size of 40 KB/layer. 
Thus, UPMEM-Precomputation has demonstrated minimal performance overhead (as low as 4\%) compared to UPMEM-Insecure.
Furthermore, as can be seen in Fig.~\ref{fig:mlp-comp}(a), the performance speedup of UPMEM-Precomputation over CPU-Insecure increases with the model size. This is due to the CPU's performance limitations under memory wall constraints, making our approach more effective when dealing with larger datasets.

As can be seen in Fig.~\ref{fig:mlp-comp}(a),  UPMEM-Precomputation-CV performs slightly better than UPMEM-Enc/Dec-V for an input size of 40 KB per layer, resulting in a $1.13\times$ speedup compared to UPMEM-Enc/Dec-V. Furthermore, increasing the input size further would lead to a more significant speedup.

Fig.~\ref{fig:mlp-comp}(b) illustrates the execution time of different MLP implementations across various batch sizes. For this evaluation, we exploit the input size of 20 KB and a weight size of 100 MB per layer. \commentone{The optimal performance, compared to CPU-Insecure and CPU-Secure, is observed at a batch size of 128, achieving an $11.63\times$ and an $11.90\times$ speedup, respectively.} Smaller batch sizes lead to underutilization of the UPMEM, 
while larger batch sizes require more PIMs to process the entire batch in parallel. Since our memory is limited to 20 PIMs, for batch sizes greater than 128, we still need to split the computation into mini-batches of 128, leading to sequential computation across different mini-batches.

Fig.~\ref{fig:mlp-comp}(a) and (b) show that the verification overhead for MLP is negligible, with an overhead as low as 12.5 ms for an input size of 40 KB per layer.

\begin{figure*}[!tp]
\centerline{\includegraphics[width=0.8\textwidth, trim={0cm 1cm 0cm 0cm}]{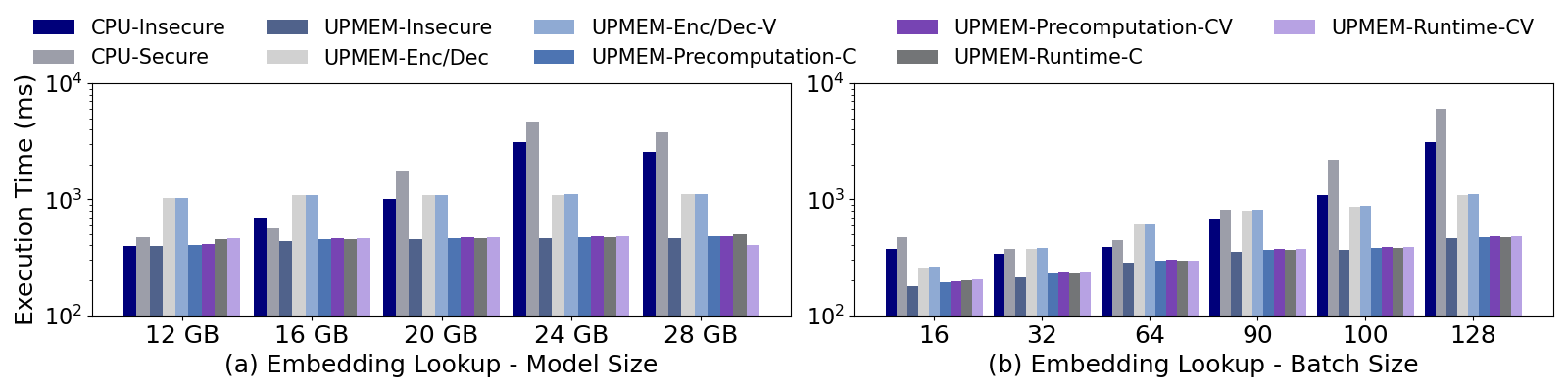}}
\caption{\commentone{Execution time comparison of different implementations of Embedding Lookup (a) with different embedding table sizes for a batch size of 128, where for larger model sizes, UPMEM-Precomputation can achieve a higher speedup compared to CPU-Secure. (b) with different batch sizes for model size = 24 GB, where UPMEM-Precomputation outperforms CPU-Secure. 
}}
\label{fig:dlrm1}
\end{figure*} 

\textbf{DLRM performance analysis:}
Fig.~\ref{fig:dlrm1}(a) demonstrates the execution time of our experiments across various embedding table sizes. We specifically report results for the embedding lookup table comparison, as this operation is the primary differentiating factor among various approaches, while the rest of the computation is performed on the CPU in all approaches.

In our performance comparison across different system configurations, we observed that our method initially exhibits a negligible speedup of $1.53\times$ for a model size of 16 GB over CPU-Insecure. However, this speedup gradually increases as the size of the embedding tables grows, eventually reaching $6.44\times$ for embedding tables of 24 GB. This trend suggests that our approach is particularly advantageous when handling larger embedding tables due to the need for irregularly accessing more considerable amounts of data.

\commentone{Our experiments show that CPU-Secure is, on average, $1.84\times$ slower than CPU-Insecure. UPMEM-Precomputation-CV achieves a $9.80\times$ speedup for embedding tables of 24 GB over CPU-Secure. 
}
Furthermore, both the precomputation and runtime schemes performed similarly in our experiments. This similarity arises because the volume of public data is relatively small, and CPU computation in the runtime scheme does not limit the overall computation.

Compared to our proposed scheme, the UPMEM-Enc/Dec approach involves additional encryption and decryption steps, which can degrade performance, mainly when processing large datasets privately. 
For example, for embedding tables of 24 GB, the overhead of decryption and encryption is about 58\% of the total computation.
 This results in a $2.30\times$ slowdown for the UPMEM-Enc/Dec approach compared to the UPMEM-Precomputation approach.

Fig.~\ref{fig:dlrm1}(b) presents the results of our embedding lookup experiments conducted across a range of batch sizes, utilizing a model size of 24 GB. Our scheme effectively accelerates the computation for larger batch sizes, while its improvement on smaller ones is insignificant. 
According to the results in Fig.~\ref{fig:dlrm1}(b), UPMEM-Insecure achieves a $6.72\times$ speedup compared to CPU-Insecure. Incorporating privacy and verification schemes leads to negligible performance degradation, and our scheme attains a $6.44\times$ speedup for a batch size of 128. This is due to the increase in irregular memory accesses as the batch size increases. However, the benefits of using PIM are less pronounced for smaller batch sizes where memory access is less significant.

\begin{figure}[htp]
\centerline{\includegraphics[width=2.8in, trim={0cm 1cm 0cm 0.8cm}]{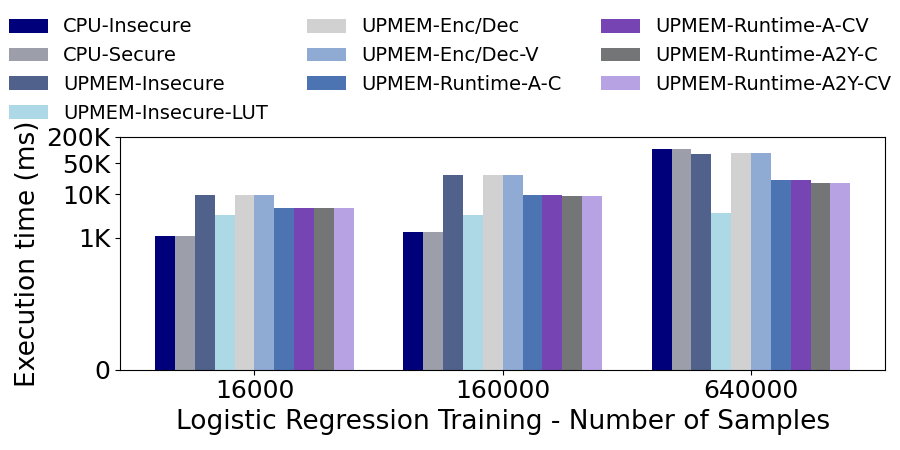}}
\caption{\commentone{Comparing different implementations of logistic regression when having different numbers of samples. UPMEM-Runtime-A(2Y) can achieve higher speedup over CPU-Secure by increasing the number of samples.}}
\label{fig:log}
\vspace{-0.3cm}
\end{figure}

\textbf{Logistic Regression performance analysis:} 
Fig.~\ref{fig:log} illustrates the execution time of different logistic regression implementations across various numbers of samples. Increasing the number of samples allows our UPMEM-Runtime scheme to achieve a speedup of up to $5.85\times$ compared to the CPU-Insecure baseline. This improvement is due to the memory wall concept.
\commentone{Our experiments on the CPU-Secure implementation of logistic regression show no overhead compared to CPU-Insecure.}

The UPMEM-Insecure baseline has different implementations for the sigmoid function of logistic regression on the UPMEM. We compared our implementation with two of their implementations: an LUT-based implementation of the sigmoid function and a non-LUT-based implementation. Our scheme performs better than the non-LUT-based implementation, but the LUT-based approach outperforms ours. This is because the non-LUT approach performs the sigmoid function on the UPMEM, which is a complex computation based on the UPMEM's capabilities. 
\commenttwo{
UPMEM-Runtime-A is implemented based on the sigmoid function; however, UPMEM-Runtime-A2Y uses the GC-friendly activation function proposed in SecureML \cite{secureml}. SecureML shows that this activation function does not degrade accuracy on their experimented datasets, or the degradation is negligible.}
For 640000 samples, our design experiences a $4.91\times$ slowdown compared to UPMEM-Insecure (LUT) but achieves a $4.61\times$ speedup compared to UPMEM-Insecure (non-LUT).

Since the verification scheme uses a linear checksum, it cannot be used to verify non-linear computations. However, we can verify the logistic regression by checking the computation twice: once after the dot product and once after the gradient descent computation. As seen in Fig.~\ref{fig:log}, the verification overhead is negligible, so performing it twice does not significantly impact the design.

As discussed in Section \ref{LinearOutsourcing}, the UPMEM-Precomputation scheme is not included here because it is not suitable for dynamic applications.

Since the TEE is trusted, Yao circuits in logistic regression can be generated without requiring any encryption. Additionally, there is a $2 \times inputs\_size$ B increase in the memory overhead and $2 \times inputs\_size$ B reduction in the communication overhead. For example, with an input size of 16000 (integer), a feature size of 16, and 1000 epochs, the switching overhead is about 846 ms, which is approximately 9\% of the total execution time. This switching results in a 32 KB increase in memory overhead, while the communication overhead between the CPU and PIM is reduced by 32 KB. 

As illustrated in Fig.~\ref{fig:log}, the arithmetic to Yao sharing approach (UPMEM-Runtime-A2Y) exhibits superior performance compared to the arithmetic sharing strategy (UPMEM-Runtime-A). This improvement stems from a reduction in the volume of data communication between the CPU and DPUs.

\begin{figure}[htp]
\centerline{\includegraphics[width=2.8in, trim={0cm 1cm 0cm 0.5cm}]{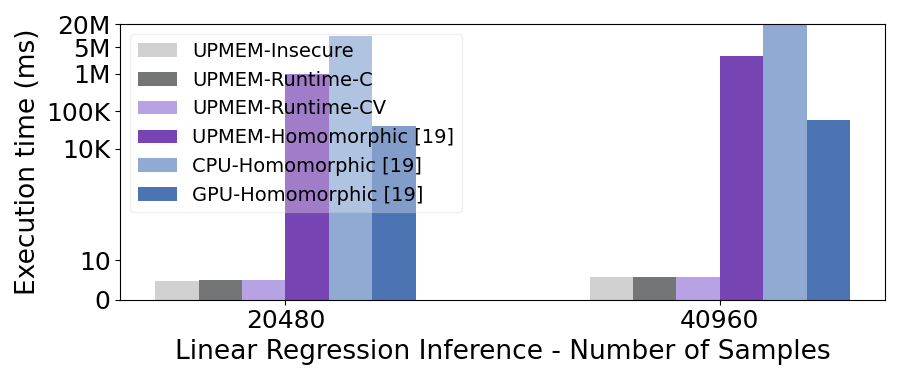}}
\caption{\ClarityText{Comparison of homomorphic-based and  MPC-based implementations of linear regression with varying numbers of samples. Our MPC-based approach consistently outperforms the HE-based method.}}
\label{fig:lin}
\vspace{-0.3cm}
\end{figure}

\textbf{Linear Regression performance analysis:} Gupta et al.~\cite{gupta2023evaluating} implement various homomorphic operations to ensure secure computation over UPMEM. Fig.~\ref{fig:lin} compares our scheme with their homomorphic-based implementation of linear regression on CPU, GPU, and UPMEM. Our scheme shows significant performance improvements over their approach, primarily due to the computational intensity of homomorphic operations. Compared to the UPMEM-Insecure, the overhead of our MPC-based approach is significantly low.

\begin{figure}[htp]
\centerline{\includegraphics[width=2.7in, trim={0cm 1cm 0cm 0.5cm}]{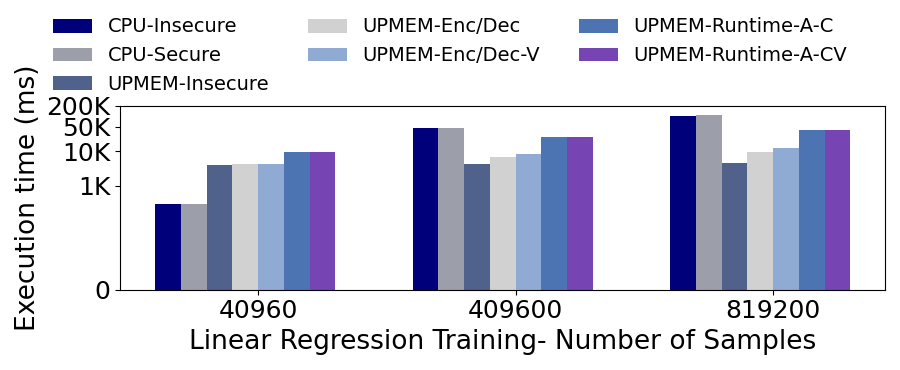}}
\vspace{-0.2cm}
\caption{\commentone{Comparing different linear regression training implementations with different numbers of samples. By increasing the number of samples, UPMEM-Runtime can achieve higher performance compared to CPU-Secure.}}
\label{fig:lin2}
\end{figure}

As illustrated in Fig.~\ref{fig:lin2}, in linear regression training, our method's performance improves as the number of samples increases compared to the CPU-Insecure approach. 
\commentone{For example, the speedup of the runtime scheme compared to CPU-Insecure and CPU-Secure is $1.73\times$ and $1.78\times$ for 409600 samples, increasing to $2.57\times$ and $2.64\times$ for 819200 samples.}
This improvement is due to memory constraints becoming more prominent with a larger number of samples. However, our method involves transferring data to the CPU mid-computation for result consolidation before returning them to UPMEM.
\commentone{Enabling TEE on the CPU for logistic regression implementation results in an average slowdown of only 2\%, which is negligible.}

Thus, our approach does not achieve the efficiency of UPMEM-Insecure due to extra inter-DPU data transfers between each iteration, resulting in a $9.20\times$ slowdown compared to UPMEM-Insecure. However, UPMEM-Runtime-CV still outperforms the CPU-Insecure.

Unlike in previous applications, UPMEM-Enc/Dec-V performs better than UPMEM-Runtime-CV in linear regression. This is because, in UPMEM-Enc/Dec-V, once the data is sent to the PIM for each iteration, no inter-DPU communication is needed. However, in both of our schemes, we need to merge the dot product results in the CPU before starting the gradient descent computation. This results in more inter-DPU communication.

\vspace{-0.3cm}
\commentsix{
\subsection{Robustness Under Threat Model} 

In this section, we discuss how our proposed framework robustly counters the defined threat model, ensuring both confidentiality and integrity in untrusted PIM architectures.
In our proposed scheme, we consider an \textit{adversary A} capable of attempting to gain unauthorized access to private data or tamper with data outside the TEE. The system's robustness under the defined threat model is evaluated as follows:

\noindent \textbf{A. Unauthorized Observation}:
\begin{itemize}[nosep]
    \item \textit{Threat}: The adversary might eavesdrop on data  stored in off-chip memory or during transmission (bus snooping). In addition, it might perform cold boot attack on either standard memory or PIM-enabled DRAM.
    \item \textit{Robustness}:
Our scheme employs counter-mode encryption and the private data is masked with OTPs. Therefore, data remains secret during transmission between memory and the CPU as well as inside the memory. Even if an adversary intercepts the data, it is indistinguishable without the secret key. 
Additionally, we are using arithmetic secret sharing and Yao's sharing for performing computations. In both schemes, the TEE holds one share of the computation, which does not reveal any information about its share of the data or the secret keys.
\end{itemize}

Security guarantees for these techniques are rooted in cryptographic principles validated by prior work (TEE~\cite{Costan2016-zb}, arithmetic secret sharing and counter-mode encryption~\cite{9773244}, and Yao's sharing~\cite{secureml}).

\noindent  \textbf{B. Unauthorized Modifications}:

\begin{itemize}[nosep]
\item \textit{Threat}: The adversary may attempt to modify the data in memory or during transmission between memory and the CPU, or might try to alter the result of the computation in the PIM. Untrusted PIMs can also seek to inject faults into the computation.

\item \textit{Robustness}: 
The MACs generated in our scheme ensure the integrity of private data by detecting any unauthorized modifications made outside the TEE. Additionally, by performing computations on the generated tags, we verify the authenticity of computations performed outside the TEE. This is achieved by comparing the retrieved results against the expected valid outputs, as described in ~\cite{9773244}.
The hashing scheme employed in our framework, based on the collision-resistant design of SecNDP~\cite{9773244}, further guarantees data integrity by ensuring that no two different inputs produce the same hash. This robust combination of MACs and collision-resistant hashing mechanisms provides a strong safeguard against data tampering and unauthorized modifications.
\end{itemize}
}

\begin{figure}[htp]
\centerline{\includegraphics[width=2.9in, trim={0cm 1cm 0cm 0.7cm}]{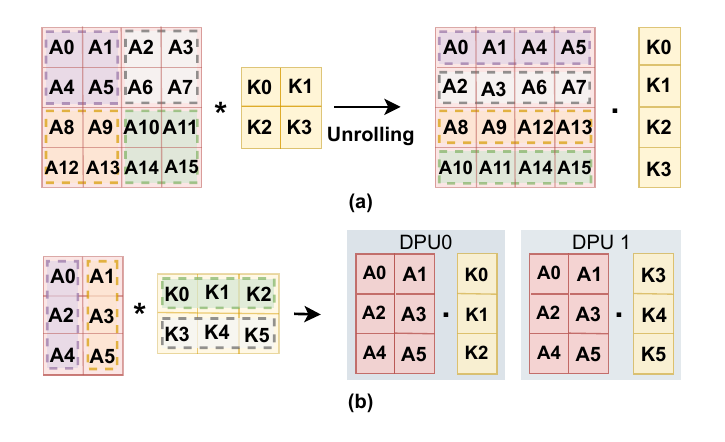}}
\vspace{-0.2cm}
\caption{\commentfive{(a) Unrolling the matrix for performing 
$2 \times 2$ convolution with stride = 2. (b) Performing GEMM on DPUs using GEMV kernel. }}
\label{unrolling}
\vspace{-0.3cm}
\end{figure}

\vspace{-3mm}
\commentfive{
\section{Expanding Real-World Applications}

Our scheme is designed to address computational bottlenecks in a variety of real-world applications by securely offloading intensive computations to PIM hardware while maintaining the necessary security guarantees. In these scenarios, PIM accelerates the bottleneck, while lighter computations are handled by the CPU under the same threat model.

In DNNs, the computational bottlenecks primarily arise from linear operations such as GEMV, GEMM, and Convolution. Our proposed secure PIM framework can be extended beyond GEMV (implemented for MLPs) to efficiently support GEMM and Convolution operations for more complex networks, including CNNs and Transformers.

For Convolution, the input matrix can be unrolled based on the kernel size and stride (Fig. \ref{unrolling}(a)), transforming the operation into GEMV computations that can be securely offloaded to PIM. Similarly, GEMM operations in Transformer models, such as in the attention mechanism, can be split into smaller GEMV tasks to fit the PIM’s parallel processing capabilities, as illustrated in Fig. \ref{unrolling}(b). For example, the computation in the attention layer of Transformers is given by:

\centering
$ \quad Attention(Q, K, V) = \text{Softmax}\left(\frac{QK^\top}{\sqrt{d_k}}\right)V$,

where the quadratic complexity of the GEMM operations (e.g., $QK^\top$) dominates the computational workload. Our framework securely partitions these GEMM tasks into GEMVs, leveraging the PIM for efficient parallel execution. If Softmax is also required on the PIM, a GC-friendly approximation of Softmax~\cite{secureml} can be utilized, allowing secure computation using Yao’s Garbled Circuits.

By demonstrating scalability across various workloads and extending to operations like GEMM and Convolution in large-scale DNNs, our framework integrates seamlessly into real-world application scenarios while maintaining robust security guarantees under the defined threat model.}

\vspace{-4mm}
\section{Conclusions}
\vspace{-1mm}
Processing-in-Memory is an approach to address the performance limitations caused by memory wall constraints. However, off-chip memory is considered untrusted based on current TEE threat models. This study introduces a low-cost method for secure computing on untrusted PIMs, which enables the offloading of computations to PIM while preserving confidentiality and allowing for the verification of the computations. We evaluated our scheme on UPMEM using data-intensive applications. The results demonstrated significant performance improvements, with up to a $14.28\times$ speedup compared to insecure CPU when offloading linear computations to the PIM, and up to a $5.85\times$ speedup when offloading both linear and nonlinear computations to the PIM. 
Our scheme not only guarantees security but also achieves the acceleration benefits of outsourcing the computation to the PIM-enabled memory.

\vspace{-0.4cm}
\section*{Acknowledgments}
This work is funded, in part, by NSF Career Award \#2339317, NSF \#2235398, the initial complement fund from UCR, and the Hellman Fellowship from the University of California. We appreciate the use of hardware resources provided by the UPMEM. We thank Prof. Nael Abu-Ghazaleh for his insightful comments. 


\section*{Ethics Considerations}

Our research focuses on securely offloading computation to an off-chip, untrusted PIM architecture using multi-party computation techniques. This research adheres to the ethical guidelines provided by USENIX Security. All authors of this paper have reviewed and fully agree with this statement. Our research does not compromise user privacy or safety, and all data used in our experiments is either synthetic or randomly generated. Therefore, no personally identifiable information (PII), sensitive data, or human subjects were involved.

We built our research upon publicly available baseline implementations. We properly acknowledged and cited the original creators and respected any licensing agreements.

Our research utilized UPMEM PIMs provided by UPMEM, which were integral to conducting our experiments. This hardware was used following the provider's guidelines and licensing agreements, and we have properly acknowledged their contribution to our research.

\section*{Compliance with Open Science Policy}

Our research adheres to the USENIX Open Science policy by making all code and scripts used in our experiments publicly available to ensure transparency and reproducibility. In the following, we outline the artifacts that will be shared:

\textbf{Description:} To ensure transparency and reproducibility, we provide all necessary source code, configuration files, and scripts. Using these resources, the performance results for UPMEM-Precompute-C(V), UPMEM-Runtime-C(V), and UPMEM-Runtime-A(2Y)-C(V) presented in Figures 13 to 18 can be reproduced.

\textbf{Access:} The code is open-source and hosted on a GitHub public repository to facilitate reuse. Additionally, it is archived on Zenodo. We encourage other researchers to explore, reproduce, and modify our code, provided they give appropriate credit.

\textbf{URL:} 
\begin{itemize}
    \item GitHub: \href{https://github.com/Secure-UPMEM/SecUPMEM.git}{github.com/Secure-UPMEM/SecUPMEM.git}.

    \item Zenodo: \href{https://zenodo.org/records/14736863}{zenodo.org/records/14736863}.

\end{itemize}
All data and models used in our experiments and performance evaluations are randomly generated.



\bibliographystyle{plain}
\bibliography{references,paperpile}


\appendix
\section{Artifact Appendix}


\subsection{Abstract}
Our artifact contains all necessary source codes and scripts for evaluating our proposed security scheme.
This paper leverages Multi-Party Computation (MPC) techniques, specifically arithmetic secret sharing, and Yao's garbled circuits, to securely outsource bandwidth-intensive computation to PIM. Additionally, we employ precomputation optimizations to prevent the CPU's portion of the MPC from becoming a bottleneck. 
We provided all the source codes and scripts for evaluating our scheme using UPMEM, the first publicly available PIM, over four data-intensive applications: Multilayer Perceptron inference (MLP), Deep Learning Recommendation Model inference (DLRM), linear regression training, and logistic regression training. 
This artifact allows researchers to reproduce our results, explore this area further, and expand our work. With this artifact, researchers can regenerate the performance results for UPMEM-Precompute-C(V), UPMEM-Runtime-C(V), and UPMEM-Runtime-A(2Y)-C(V), as presented in figures 13 to 18.
\subsection{Description \& Requirements}


\subsubsection{Security, privacy, and ethical concerns}
Our research does not compromise user privacy or safety, and all data used in our experiments is either synthetic or randomly generated. Therefore, no Personally Identifiable Information (PII), sensitive data, or human subjects were involved.

\subsubsection{How to access}
The code is open-source and hosted on a GitHub public repository to facilitate reuse. Additionally, it is archived on Zenodo. We encourage other researchers to explore, reproduce, and modify our code, provided they give appropriate credit. We always recommend using the latest Zenodo release via the provided concept DOI.

\begin{itemize}[nosep]
    \item Zenodo: \href{https://zenodo.org/records/14736863}{zenodo.org/records/14736863}.

     \item GitHub: \href{https://github.com/Secure-UPMEM/SecUPMEM.git}{github.com/Secure-UPMEM/SecUPMEM.git}.
     
\end{itemize}

\subsubsection{Hardware dependencies}
To evaluate our method, we utilize UPMEM PIM hardware, which primarily consists of standard DDR4-2400 DIMMs integrated with DPUs. Our setup includes 20 PIM-enabled DIMMs, providing a total of 160 GB of MRAM and 2560 DPUs working in parallel at a clock frequency of 350 MHz. The host server for the UPMEM system is equipped with a 2-socket Intel Xeon Silver 4110 CPU. To accurately reproduce our results, access to the actual hardware is necessary. UPMEM’s PIM data centers are accessible upon request at \url{https://www.upmem.com/developer.}

\subsubsection{Software dependencies}\label{software}
All the implemented applications require the UPMEM SDK, which can be installed based on the hardware specifications and is accessible at \url{https://sdk.upmem.com.} 
For dependencies related to MLP, DLRM, logistic regression, and linear regression, please refer to our baseline implementations as our work is built upon them.
\begin{itemize}[nosep]
    \item MLP \href{https://github.com/CMU-SAFARI/prim-benchmarks} {(Link)}
    \item DLRM \href{https://github.com/UBC-ECE-Sasha/PIM-Embedding-Lookup}{ (Link 1,}
    \href{https://github.com/upmem/PIM-Embedding-Lookup/tree/multicol/upmem}{ Link 2)} 
    \item Logistic and Linear Regression \href{https://github.com/CMU-SAFARI/pim-ml}{ (Link 1,} \href{https://github.com/CMU-SAFARI/prim-benchmarks}{ Link 2)} 
\end{itemize}
\vspace{3mm}

\subsubsection{Benchmarks}

Our scheme is evaluated using MLP inference, DLRM inference, logistic regression training, and linear regression training.
To evaluate our implementation, we use randomly generated inputs.
\subsection{Set-up}

\subsubsection{Installation}
To run this artifact on your local device, the UPMEM SDK must first be installed, which is available at \url{https://sdk.upmem.com}. However, as previously mentioned, reproducing our results requires access to real hardware (20 UPMEM PIMs) rather than the simulator.
Once the UPMEM SDK is installed, the artifact can be downloaded from \href{https://github.com/Secure-UPMEM/SecUPMEM.git}{GitHub} and \href{https://zenodo.org/records/14736863}{Zenodo}.


\subsubsection{Basic Test}

Script \textit{run\_functionality.sh}, in the root directory, can be used to perform a simple functionality check. This file executes a basic test on all the applications sequentially and outputs execution time. To run a specific application, the \textit{run\_functionality.sh} script in the corresponding folder can be executed.


\subsection{Evaluation workflow}

\subsubsection{Major Claims}
Our major claim is as follows:

\begin{compactdesc}


    \item[(C1):] 
Compared to a secure CPU implementation, our framework achieves speedups of $14.66\times$, $9.80\times$, $2.64\times$, and $5.85\times$ for MLP inference, DLRM inference, Linear Regression training, and Logistic Regression training, respectively. This is proven by the experiment (E1) described in Section 7.1 whose results are illustrated in Figures 14, 15, 16, 18.

\end{compactdesc}

\subsubsection{Experiments}

\begin{compactdesc}







 \item[(E1):]
 [30 human-minutes + 2.5 compute-hour + 32GB disk]:

    \begin{asparadesc}
        \item[How to:] Our results can be reproduced by following the three steps below.

        \item[Preparation:] 
Install the UPMEM SDK, as described in Section \ref{software}, then clone our artifact.
        

        \item[Execution:]
        Script \textit{./run\_reproduce.sh}, in the root directory, can be used to regenerate our results. This file executes all the applications with our configuration sequentially and outputs execution time for UPMEM-Precompute-C(V), UPMEM-Runtime-C(V), and UPMEM-Runtime-A(2Y)-C(V), as presented in Figures 13 to 18. To reproduce the results for a specific application, the \textit{./run\_reproduce.sh} script in the corresponding folder can be executed.
        

        This script compiles and links the host and DPU source codes for a specific number of DPUs and tasklets. There is a Makefile for each application that facilitates the compilation and linking of the source codes.  
        
        
        \item[Results:]  After running our script, the final execution time is reported. Table \ref{tab:notations} explains the different notations used for manually determining execution times. The execution time is calculated using the formula provided below:    
\[
PIM\ Time = CPU-DPU + PIM\ kernel + DPU-CPU
\]
\[
Kernel\ time = Max[CPU\ Time, PIM\ Time]
\]
\[
Execution\ time = Kernel\ time + Merge + Verification
\]

\begin{table}[htp]
	\centering
	\caption{List of timing notations}
	\scalebox{0.80}{
	\begin{tabular}{>{\centering\arraybackslash}p{1in}>{\centering\arraybackslash}p{2.5in}}
		\toprule
  Timing notations & Explanation\\
		\midrule
    $CPU-DPU$& CPU to DPUs transfer time \vspace{1mm}\\
    $DPU-CPU$ & DPUs to CPU transfer time \vspace{1mm}\\
    $ PIM\ kernel$& Kernel execution time on DPUs \vspace{1mm}\\
     $CPU\ Time$ & Kernel execution time on CPU \vspace{1mm}\\
     $Kernel\ time$ & The maximum of CPU Time and PIM Time \vspace{1mm}\\
     $Merge$ & Merging time of CPU and PIM results \vspace{1mm}\\
     $Verification$ & Verification time \vspace{1mm}\\
		\bottomrule
	\end{tabular}}
	\label{tab:notations}
 \vspace{0.2cm}
\end{table}
    \end{asparadesc}
\end{compactdesc}




\subsection{Version}
Based on the LaTeX template for Artifact Evaluation V20231005. Submission,
reviewing and badging methodology followed for the evaluation of this artifact
can be found at \url{https://secartifacts.github.io/usenixsec2025/}.

\end{document}